# More is Different: Mobile Ions Improve the Design Tolerances of Perovskite Solar Cells

Hart, L. J. F.[*], Angus, F.J., Li, Y., Khaleed, A., Durrant, J. R., Djurišić, A. B., Docampo, P.[*], and Barnes, P. R. F.[*]

L. J. F. Hart, Dr P. R. F Barnes

Department of Physics and Centre for Processable Electronics, Imperial College London, South Kensington, U.K.

L. J. F. Hart, Professor J. R. Durrant

Department of Chemistry and Centre for Processable Electronics, Imperial College London, 82 Wood Lane, London, U.K.

F. J. Angus, Dr P. Docampo

Department of Chemistry, University of Glasgow, University Avenue, Glasgow, U.K.

Professor A. B. Djurišić, Y. Li, A. Khaleed

Department of Physics, The University of Hong Kong, Pokfulam Road, Hong Kong S.A.R, China

Professor J. R. Durrant

SPECIFIC IKC, College of Engineering, Swansea University, Bay Campus, Fabian Way, Swansea, U.K.

Email: lucy.hart18@imperial.ac.uk; pablo.docampo@glasgow.ac.uk; piers.barnes@imperial.ac.uk

## Abstract

Many recent advances in metal halide perovskite solar cell (PSC) performance are attributed to surface treatments which passivate interfacial trap states, minimise charge recombination and boost photovoltages. Surprisingly, these photovoltages exceed the cells' built-in potentials, often with large energetic offsets reported between the perovskite and transport layer semiconductor band edges – contradicting standard photovoltaic design principles. Here we show that this tolerance to energetic offsets results from mixed ionic/electronic conduction in the perovskite layer. Combining drift-diffusion simulations with experiments probing the current-voltage performance of PSCs as a function of ion distribution, we demonstrate that electrostatic redistribution of ionic charge reduces surface recombination currents at steady-state, increasing the photovoltage by tens to hundreds of millivolts. Thus, mobile ions can reduce the sensitivity of photovoltage to energetic misalignments at perovskite/transport layer interfaces, benefitting overall efficiency. Building on these insights, we show how photovoltaic design principles are modified to account for mobile ions.

## 1. Introduction

Metal halide perovskites are a promising class of materials used as the active layer in next-generation photovoltaics. The highest performing, lab-scale, perovskite solar cells (PSCs) have achieved efficiencies of over 25%,[1,2] close to the best silicon photovoltaics. However, their underlying device physics is fundamentally different due to high densities of mobile halide vacancies in the perovskite layer.[3–9] The slow redistribution of this ionic charge to screen internal electric fields in the perovskite devices explains the well documented hysteresis in their current-voltage characteristics and also leads to extraction losses of photogenerated charge carriers at short circuit operating conditions.[10–12] Consequently, mobile ions are typically considered to be detrimental to device performance, even without taking into account their potential role in cell degradation.[13,14] However, the redistribution of mobile ions also influences recombination at operating voltages close to open circuit ($V_{OC}$) which impacts the steady-state power conversion efficiency (PCE).

Halide vacancies have been shown to form shallow defects and thus their presence does not limit bulk electronic carrier lifetimes.[15,16] Consequently, their impact on $V_{OC}$ and efficiency will primarily depend upon how they affect device electrostatics and hence the electronic carrier distribution. Recombination at interfaces is thought to limit $V_{OC}$ in even the best PSCs. It has been suggested that accumulation of ionic charge at perovskite/transport layer interfaces could reduce surface recombination currents by repelling minority electronic carriers, similar to the mechanism of field effect passivation used in silicon photovoltaics.[8,17–22] Indeed, recent simulations have shown that ion redistribution can significantly increase $V_{OC}$, and thus PCE, in situations where the built-in potential of the device falls below $V_{OC}$.[22] This is relevant to most PSCs, particularly high-efficiency devices, where values of $V_{OC}$ routinely surpass the built-in potential (typically reported to lie in the range 0.8-1.0 V).[23–25]

Currently, there is no direct experimental confirmation of this effect, and previous experiments have been interpreted as implying that mobile ions have no significant impact on steady-state $V_{OC}$.[11,13] Here, we address this apparent contradiction by proposing a novel approach to accurately assess the impact of mobile ions on PSC performance. Using this method, we measure the change in $V_{OC}$ due to mobile ions for a variety of device architectures and perovskite compositions. In all cases, we find that mobile ions do increase $V_{OC}$, even in high efficiency devices showing negligible hysteresis. Additionally, the experiments are in excellent agreement with the predictions of our drift-diffusion simulations. We use these simulations to demonstrate that, in all cases of interest, the presence of mobile ions makes device performance more tolerant to energetic offsets between the conduction/valence band edges of the perovskite and its transport layers, at the cost of an increased sensitivity to the rate of interfacial recombination. This means that, although mobile ions do not *a priori* reduce maximum achievable PCEs, they do change the choice of optimal transport layer materials and we show how standard photovoltaic design principles should be modified to account for ionic redistribution.

## 2. Results

### 2.1 Theory and Simulation

Field effect passivation is a technique commonly applied in silicon photovoltaics and involves the insertion of a layer of static ionic charge between the silicon active layer and its transport layers.[26] This ionic charge serves to repel minority carriers and thus reduce rates of surface recombination, improving photovoltage. It has been suggested that the presence of mobile ionic charge in PSCs should have similar consequences for steady state device performance.[20–22] We illustrate this in **Figure 1a**, which shows the impact of mobile ions on solar cell band diagrams for applied voltages below (top row), equal to (centre) and above (bottom row) the flat band condition across the active layer, $V_{flat}$.

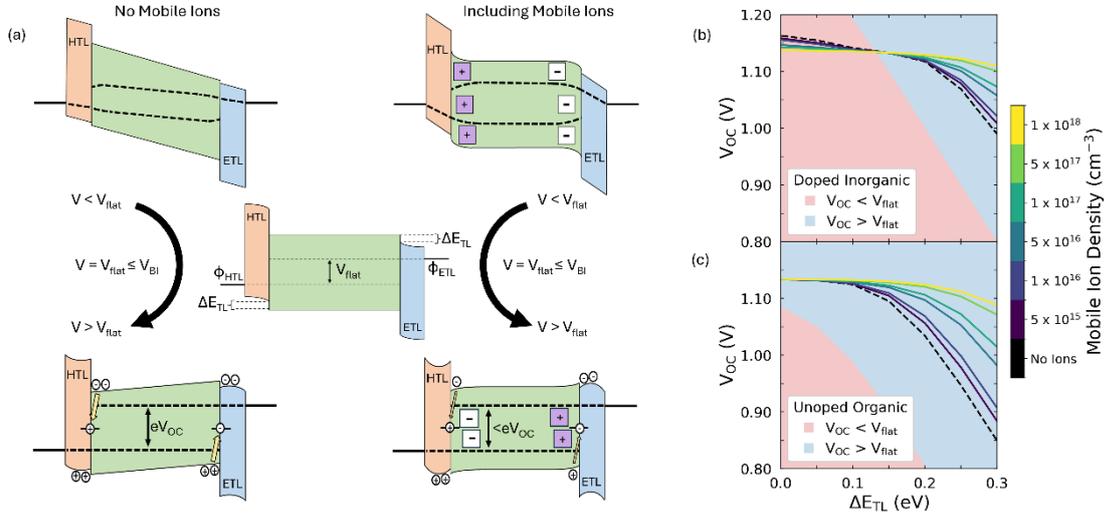

**Figure 1**: (a) Illustration of the impact of mobile ions on solar cell band diagrams. In these figures, we assume undoped organic transport layers. $\Delta E_{TL}$ is defined as the energetic offset between the perovskite's conduction (valence) band and that of the ETL (HTL). The built-in potential ($V_{BI}$) is defined at the difference in the equilibrium values of the anode and cathode work function ($\varphi_{HTL}$ and $\varphi_{HTL}$, respectively). In the central panel, where the applied voltage is equal to the flat band condition in the active layer, $V_{flat}$, the band diagrams of devices with and without mobile ions look identical. Note how the residual band bending in the undoped contacts means that $V_{flat} < V_{BI}$ in this example. In the bottom row, we illustrate how band bending due to the redistribution of mobile ions can reduce minority carrier accumulation at interfaces, when compared to an equivalent device with no ions. This reduces the rate of interfacial recombination for a given applied voltage, increasing the device's open circuit voltage ($V_{OC}$). Figures (b) and (c) show the dependence of $V_{OC}$ on $\Delta E_{TL}$ for doped inorganic and undoped organic parameter sets, respectively. The red and blue shaded regions indicate where $V_{OC}$ is less than and greater than $V_{flat}$, respectively.

The presence of mobile ions increases the capacitance of the perovskite leading to a smaller change in the voltage across the active layer and greater changes in voltage across the transport layers (see **Figures S1**). This means that, for a given applied voltage above (below) $V_{flat}$, the magnitude of accumulation (depletion) of minority carriers at the perovskite/transport layer interfaces will be greater when no mobile ionic charge is available to electrostatically substitute for the electronic carriers. Thus, the presence of mobile ions will reduce surface recombination currents at applied voltages greater than $V_{flat}$ as long as the accumulation/depletion of ionic charge at interfaces does not lead to a degradation of the transport layers with a corresponding increase in interfacial recombination rate. This condition is applicable when degradation has not yet occurred or where interfaces have been well passivated so that degradation is prevented. We note that the latter criteria must apply for any commercially viable PSC technology, meaning that the impact of mobile ions on steady state performance will only become more important as passivation strategies improve.

To systematically explore how the choice of transport layers affects $V_{flat}$ and thus the influence of ions on $V_{OC}$, we simulated JV scans on three-layer, symmetric PSCs (device structure illustrated in **Figure 1a**) and contrasted the behaviour with and without the inclusion of a mobile ionic species for two different transport layer parameter sets (see **Methods** for full details of the simulation protocol and assumptions made in device modelling). In the 'doped inorganic' parameter set, we considered materials which have a high intrinsic carrier density, mobility, and permittivity. These properties are representative of n-type metal oxides, such as $SnO_2$,[27–29] and doped Spiro-OMeTAD,[30,31] though the latter, being an organic material, does not have a high permittivity. In the 'undoped organic' parameter set, we considered materials which are intrinsic semiconductors, with a low mobility and permittivity. These properties reflect organic semiconductors which are commonly used as the transport layers in p-i-n PSCs (e.g., $C_{60}$, PCBM, PTAA).[32,33] The parameters of the transport layers are summarised in **Table S1** and all other parameters are given in **Table S2**.

When choosing transport layer materials, two properties are commonly seen as having the largest impact on device performance: transport layer energetic offset, $\Delta E_{TL}$ (as defined in **Figure 1a**) and surface recombination velocity, $v_S$. The latter parameter determines the rate of interfacial recombination at the perovskite/transport layer interfaces and is important because this process is thought to limit $V_{OC}$ in the majority of the highest performing PSCs, especially those which use $C_{60}$ as the ETL.[34–36] As $\Delta E_{TL}$ and $v_S$ have the potential to be easily tuned by material selection [12] and the use of interfacial passivation strategies,[32] respectively, we investigated the effect of mobile ions on $V_{OC}$ over an experimentally relevant range of these parameters. In the main text, we consider the cases where both parameters are varied symmetrically as this simplifies the description while maintaining most of the relevant device physics. However, special care must be taken when there is significant difference between the values of $\Delta E_{TL}$ at the perovskite/ETL and perovskite/HTL interfaces as there will not be a single voltage at which the ion population is uniformly distributed in such cases. This is because the accumulation and depletion regions on each side of the perovskite invert at different applied potentials due to charge carrier imbalances in the active layer.[19] We provide a brief discussion of such cases in **Supplementary Note One**, and a comprehensive mathematical treatment can be found in Cordoba and Taretto.[22]

Our results for varying $\Delta E_{TL}$ are shown in **Figure 1b** and **Figure 1c** for the doped inorganic and undoped organic parameter sets, respectively. We observe that the presence of mobile ions can increase $V_{OC}$ for both parameter sets and that the size of this effect increases with both $\Delta E_{TL}$ and the mobile ionic density. A result of this is that a high mobile ion density makes the value of $V_{OC}$ less dependent upon the energetic alignment between the perovskite and the transport layers. These observations are natural consequences of the mechanism shown in **Figure 1a**. First, the improvement in $V_{OC}$ due to mobile ions is greater in a device with a higher mobile ion density as there is more ionic charge available to electrostatically compensate minority electronic charge at the interfaces with the transport layers. Secondly, ions are more beneficial to $V_{OC}$ in devices with larger values of $\Delta E_{TL}$ as large values of $\Delta E_{TL}$ restrict $V_{BI}$, resulting in lower values of $V_{flat}$ (see **Figure S2**). Only once $V_{OC}$ lies above $V_{flat}$, do ions become beneficial to $V_{OC}$ (compare the impact of mobile ions in the red and blue regions of **Figures 1b-c**) as they can mitigate the high rates of surface recombination which are present in devices where there is a reverse field in the active layer and no mobile ions. Thus, ions facilitate a decoupling of $V_{OC}$ and $V_{flat}$.

**Figures 1b-c** also demonstrate that ions result in a larger increase in $V_{OC}$ in the case of devices with undoped, organic transport layers. This is because these transport layers result in lower $V_{flat}$ values for a given $\Delta E_{TL}$, despite both sets of simulations using the same values of $V_{BI}$. We can understand this by considering what determines $V_{flat}$ in each configuration. In the doped, inorganic case, we find that $V_{flat}$ is dictated by the offset in the equilibrium values of the quasi-Fermi levels of the transport layers (although we note that it falls below this value when there are significant injection barriers from the electrodes into the transport layers, as shown in **Figure S2**). In contrast, for symmetric, undoped interlayers, $V_{flat}$ can be found by numerically solving the implicit equation (see **Supplementary Note Two**)

$$w_{TL} = \sqrt{\frac{2\varepsilon\varepsilon_0 k_B T}{n_0 e^2}} \exp\left(\frac{e(V_{BI}-V_{flat})}{4k_B T}\right)\left[\frac{\pi}{2} - \arcsin\left(\exp\left(\frac{-e(V_{BI}-V_{flat})}{4k_B T}\right)\right)\right] \tag{1}$$

where $w_{TL}$ is the transport layer width, $\varepsilon_0$ the permittivity of free space, $\varepsilon$ the transport layer permittivity, $k_B$ the Boltzmann constant, $T$ the temperature and $n_0$ the carrier density in the transport layer at the interface with the electrode, which depends upon the transport layer density of states and the offset between the electrode work function, $\varphi$, and the relevant band edge of the transport layer (i.e., for the HTL, $n_0 = N_V \exp[(E_V - \varphi)/k_B T]$. We note that this equation only depends upon transport layer properties, which implies that using undoped interlayers can result in a substantial reduction in the value of $V_{flat}$, regardless of the presence of mobile ions. However, when mobile ions are present, they can compensate for this reduction in $V_{flat}$ and allow devices with undoped transport layers to achieve $V_{OC}$ values almost as high as their doped counterparts.

In addition to these positive impacts, varying $v_S$ allowed us to identify two regimes where the presence of mobile ions does not improve $V_{OC}$ (see **Figures S3-4**). First, mobile ions will not improve $V_{OC}$ when $V_{OC}$ is less than $V_{flat}$ since then the voltage drops in the perovskite layer are positive. In this situation, the presence of mobile ions results in a higher minority carrier density at the transport layer interfaces than is the case without mobile ions, and so surface recombination losses are greater for the same applied voltage, reducing $V_{OC}$. Secondly, for devices operating in the high injection limit, mobile ions do not improve $V_{OC}$ when recombination losses are dominated by bulk processes. This is because, assuming midgap trap states and equal carrier lifetimes, the rate of Shockley-Read-Hall recombination is maximised when the electronic carrier densities are equal (i.e., $n \approx p$). This is true in a greater fraction of the perovskite layer when mobile ions are present as they screen the electric field, which results in more uniform carrier distributions and thus a greater overlap of the electrons and hole populations.

## 2.2 Experimental Validation of the Impact of Mobile Ions on Open-Circuit Voltage

With these theoretical expectations in hand, we now seek to test the predictions of our drift-diffusion simulations experimentally. To do this, we make use of the Stabilise and Pulse technique described by Hill et al.,[37] with full details of the measurement protocol given in the **Methods** section. In brief, this technique allows one to measure JV curves for PSCs with the ions frozen in a configuration determined by the stabilisation voltage, $V_{bias}$. This is achieved by first holding the PSC at $V_{bias}$ for a stabilisation period, during which the mobile ions reach quasi-steady state (QSS) for the chosen bias condition. Then, the current output is measured for a range of short pulse voltages superimposed on $V_{bias}$. By plotting the current during the pulses versus the pulse voltages, a JV curve can be reconstructed for the steady state ion distribution defined by $V_{bias}$ since there will be negligible ionic redistribution if the pulses are sufficiently short and the duty cycle small enough, i.e., < 10%.[20,37–39] If $V_{bias}$ is chosen to equal $V_{flat}$ and assuming an energetically symmetric device, the reconstructed JV curve will be the one obtained with the ions distributed uniformly across the active layer, and thus equivalent to the JV curve which would be obtained for the same device structure, but without a mobile ionic species.

To estimate $V_{flat}$, we performed Stabilise and Pulse measurements at multiple values of $V_{bias}$ and analysed the change in gradient of the JVs around $V_{OC}$, as detailed in ref. [37] (further details given in the **Methods**). As noted above, in situations where there are asymmetric energetic offsets to the transport layers, there will not be a single value of $V_{flat}$. Thus, we have performed additional simulations to verify that the Stabilise and Pulse technique can still extract the 'ion-free' device performance in such cases, and these are discussed in **Supplementary Note One**. Lastly, by plotting the average currents from the last 30 seconds of the stabilisation period versus $V_{bias}$ we can determine the JV curve with the ions at QSS (see **Figure S5**).[40,41] By combining the QSS JV and the reconstructed Stabilise and Pulse JV (SaP JV) evaluated at $V_{bias} \approx V_{flat}$, we can experimentally determine the JV curves of a given device with and without a mobile ionic species to evaluate the effect of ions on device performance.

Based on our simulations, we hypothesise that we will see a smaller impact of mobile ions on $V_{OC}$ if we reduce the fraction of the total recombination which occurs at perovskite/transport layer interfaces (see **Figures S2-3**). To test this experimentally, we measured the change in $V_{OC}$ due to ions in n-i-p PSCs with and without a self-assembled monolayer (SAM) between the perovskite and the ETL. The device stack was Au/Spiro-OMeTAD/MAPbI3/(SAM)/TiO2/FTO (see **Figure S6**), where methyl ammonium lead iodide (MAPI) was chosen since, although it limits our maximum device efficiency, it has been widely reported to have a high mobile ion density ($10^{17}$-$10^{19}$ cm$^{-3}$).[42–45] Our simulations show that this will maximise any effect of mobile ions on $V_{OC}$, making such an effect easy to detect experimentally. For our SAM, we used the benzoic acid derivative of C60 (C60-BA) which has been shown to enhance the efficiency of electron extraction and to passivate shallow trap states at the perovskite/TiO2 interface.[46–48] Device parameters with and without C60-BA are summarised in **Tables S3-4**.

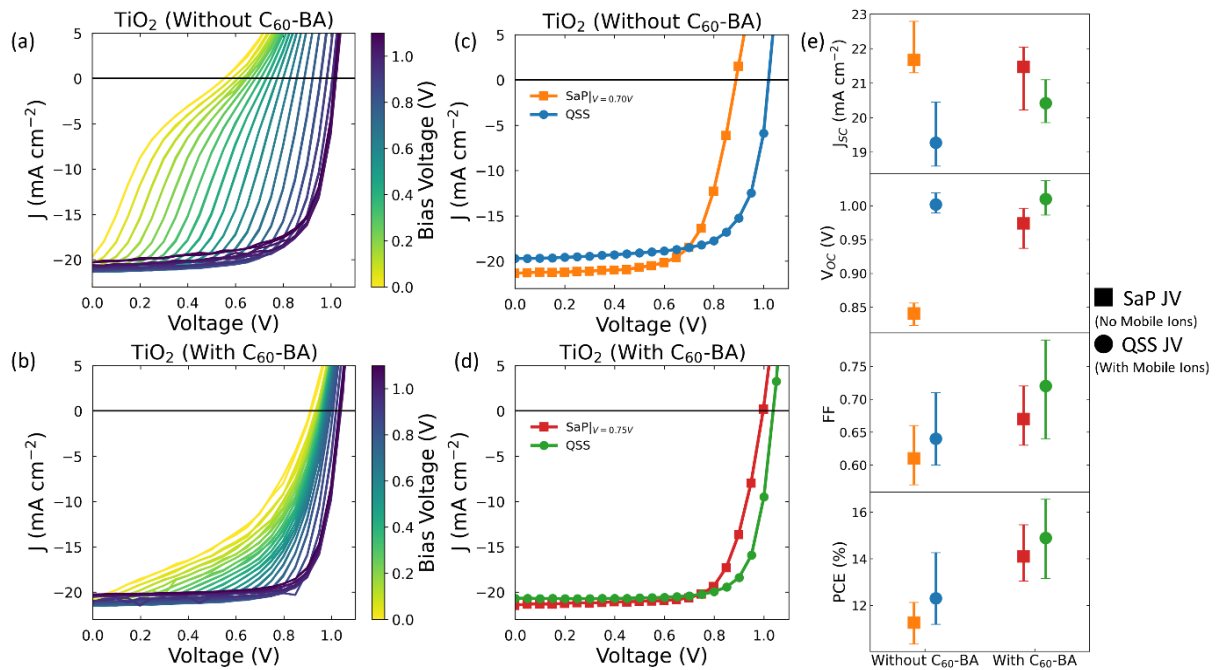

**Figure 2:** Stabilise and Pulse measurement for the devices (a) without and (b) with C60-BA. The quasi-steady state (QSS) JVs and Stabilise and Pulse (SaP) JVs evaluated at $V \approx V_{flat}$ for the TiO2 devices (c) without and (d) with C60-BA. (e) Summary of the JV parameters extracted from the SaP JVs and QSS JVs for all measured devices. Error bars indicate the range of measured values.

In **Figure 2a** and **Figure 2b**, we show the results of Stabilise and Pulse measurements performed on devices without and with C60-BA, respectively (relevant solar simulator data is shown in **Figure S7**). The Stabilise and Pulse data were analysed to extract values for $V_{flat}$ as shown in **Figure S8**. When averaged over all measured devices, $V_{flat}$ was found to be $0.66 \pm 0.02$ V for the devices without C60-BA and $0.73 \pm 0.02$ V for those with C60-BA (for details of all measured $V_{flat}$ values, see **Tables S3-4**). We note that, although our values of $V_{flat}$ are low compared to some others in the literature, many literature values were obtained from Mott-Shockley analysis, the reliability of which has been challenged when applied to PSCs.[49–51]

Next, we use our values of $V_{flat}$ to identify the SaP JV with the most uniform mobile ion distribution and consider this to be an experimental measure of device performance in the absence of mobile ions. This allows us to plot experimentally obtained JVs which illustrate how device performance is altered by the presence of mobile ions. We show these in **Figure 2c** and **Figure 2d** for the cases without and with $C_{60}$-BA, respectively. These plots demonstrate that the presence of mobile ions increases the $V_{OC}$ of both devices (see also **Table 1**). When evaluated at $V_{bias} \approx V_{flat}$, the average of these increases are $160 \pm 10$ mV and $40 \pm 10$ mV for the devices without and with $C_{60}$-BA, respectively (see **Figure S9** for how this difference varies with $V_{bias}$). We see that the size of the $V_{OC}$ improvement is smaller for the device with $C_{60}$-BA, which is consistent with our hypothesis. However, this hypothesis was based on simulations which used idealised, symmetric devices. To verify that the insights from this simplified model apply to real PSCs, we performed explicit simulations of the MAPI devices shown in **Figure 2** in which we assumed that the effect of the $C_{60}$-BA was to minimise surface recombination at the perovskite/ETL interface. We obtained an excellent agreement between the simulated and measured trends (see **Table 1** and **Supplementary Note Three**), demonstrating our model's applicability to complex experimental situations.

| Device | Technique | $V_{OC}$ with No Mobile Ions (V) | $V_{OC}$ with Mobile Ions (V) | $\Delta V_{OC}$ Due to Ions (mV) |
|---|---|---|---|---|
| Without $C_{60}$-BA | Simulation | 0.88 | 0.99 | 110 |
| | Experimental | $0.84 \pm 0.02$ | $1.00 \pm 0.01$ | $160 \pm 10$ |
| With $C_{60}$-BA | Simulation | 0.95 | 1.00 | 50 |
| | Experimental | $0.97 \pm 0.02$ | $1.01 \pm 0.02$ | $40 \pm 10$ |

**Table 1:** Summary of the experimental and simulated $V_{OC}$ values obtained for the Au/Spiro-OMeTAD/MAPI/TiO$_2$/FTO device structure, with and without the inclusion of $C_{60}$-BA.

In addition to having a higher $V_{OC}$, **Figure 2e** shows that the mean PCE of the QSS JVs is higher than that of the corresponding 'no mobile ions' SaP JVs evaluated at $V_{flat}$. This is even more striking when we consider the device-by-device data listed in **Tables S3-4**, which reveal that the same trend is found in every measured device bar one, where performance was comparable in the SaP and QSS JVs. To confirm that this effect is not limited to devices using TiO$_2$ as the ETL, we performed the same measurements on devices with $SnO_2$ in the place of $TiO_2$ and observed similar trends (see **Supplementary Note Four**).

Thus, our results demonstrate that the presence of mobile ions can improve device performance in two commonly used device architectures. We note here that our results appear to contradict previous studies which compared fast and slow JV sweeps to assess PSC performance with and without mobile ions and observed no significant impact of ions on $V_{OC}$.[11,13] However, in these experiments, the device performance without mobile ions was assessed following a stabilisation period under illumination at open circuit. Consequently, $V_{OC}$ was predetermined to be identical for both the slow and fast JV measurements since this is the only applied voltage in which the device configuration is identical in both scans. As a result, this measurement protocol gives no information about whether the distribution of ions at steady state influences $V_{OC}$ relative to an equivalent device without mobile ions except in the special case where $V_{flat} = V_{OC}$, which is not generally applicable.

## 2.3. Impact of Mobile Ions on Hysteresis-Free, High Efficiency Perovskite Solar Cells

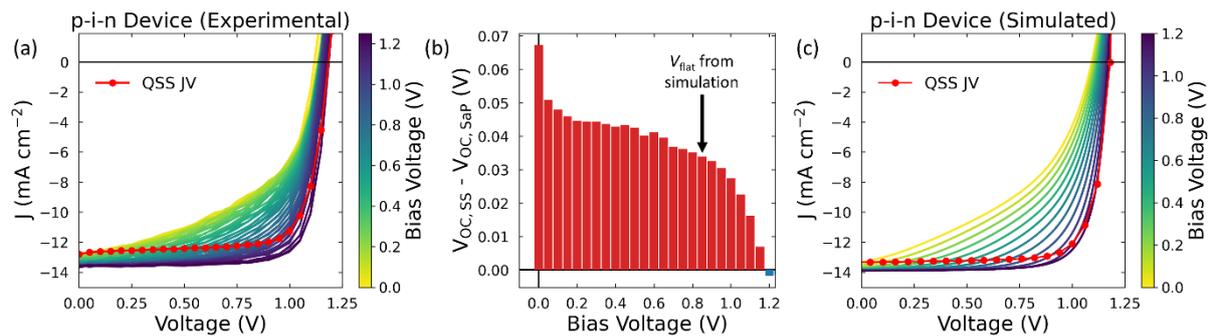

**Figure 3:** (a) Results of the Stabilise and Pulse measurement for a high efficiency p-i-n device (device structure shown in **Figure S6**). A light intensity of ~0.5 Suns was used to prevent device degradation during the measurement (see **Methods**). The red line indicates the quasi-steady state (QSS) JV, in which the ions are always in their dynamic-equilibrium distribution for the applied voltage. (b) Difference between the $V_{OC}$ of the QSS JV and the stabilise and pulse (SaP) JVs for each applied prebias. The black arrow indicates the value of $V_{flat}$ extracted from our simulations of this device. (c) Simulated Stabilise and Pulse data using parameters representative of the device stack shown in (a). By examining the simulated quasi-steady state JV, we identified that the ionic distribution was uniform at an applied bias of 0.85 V (see **Figure S13**). We note that this is less than the work function offset between the cathode and anode given in **Table S5** (1.05 V), which is due to the undoped nature of the ETL (see **Equation 1** and **Supplementary Note Two**).

Until now, we have focused on the pronounced effects of mobile ions in devices using MAPI as the active layer. However, the highest efficiency devices commonly use more complex perovskite compositions which have been reported to have lower concentrations of mobile ions.[42] Our simulations predict that the change in $V_{OC}$ due to mobile ions decreases as the mobile ion concentration decreases (see **Figures 1b-c**). This raises the question: are mobile ions relevant to the steady-state performance of the highest efficiency PSCs?

To address this question, we performed Stabilise and Pulse measurements on high efficiency, p-i-n devices which showed little JV curve hysteresis when measured at scans rates in the range $0.01 - 0.50$ Vs$^{-1}$ (PCE of 21.5% for the device shown in the text, see **Figure S10** for JV data) with the structure Ag/BCP/PCBM/PEAI/perovskite/2PaCz/NiO$_x$/ITO where the perovskite composition is Cs$_{0.05}$(FA$_{0.87}$MA$_{0.13}$)Pb(I$_{0.87}$Br$_{0.13}$)$_3$ (see **Figure S6** for the device stack and **Figure S11** for the stabilisation data). The full results are shown in **Figure 3a**, where we have superimposed the QSS JV to allow for a direct comparison with the pulsed JV data. We have done this to illustrate the significant effect that the mobile ion distribution has on the shape of the JV curve, even in this "hysteresis-free" device. We also quantify the effect of mobile ions on $V_{OC}$ as a function of $V_{bias}$ (i.e., different 'frozen ion' distributions) by calculating the difference between $V_{OC}$ at quasi steady state ($V_{OC,SS}$) and that determined from the SaP JVs measurements for each value of $V_{bias}$ ($V_{OC,SaP}$). This is shown in **Figure 3b**, which demonstrates that the presence of ions improves $V_{OC}$ for $V_{bias} < 1.20$ V (i.e., until $V_{bias} \approx V_{OC,SS}$) and improves PCE for $V_{bias} < 1.00$ V (see **Figure S12**).

Due to the small change in the gradient of the SaP JVs around $V_{OC}$, we did not rely solely on our experimental analysis to extract the value of $V_{flat}$ for this device. We also performed additional drift-diffusion simulations to identify this quantity using the parameter set listed in **Table S5**. As is shown in **Figure 3c**, we could use these parameters to simulate Stabilise and Pulse measurements which showed a good quantitative agreement with the experimental data. This gives us confidence in the accuracy of our input parameters and thus the simulated $V_{flat}$ value of 0.85 V (see **Figure S13**). We note that this value is comparable to that extracted from our experimental data ($V_{flat} = 0.83$ V, see **Figure S14**) and to literature values for similar device stacks.[52] When we evaluate $V_{OC,SS}$ - $V_{OC,SaP}$ at this value of $V_{flat}$ (indicated by the arrow on **Figure 3b**), we find that the presence of mobile ions increases $V_{OC}$ by approximately thirty millivolts in this class of high efficiency p-i-n devices, demonstrating that mobile ions remain relevant to the device physics of high performance PSCs.

### 2.4 Impact of Mobile Ions on Design Rules for Perovskite Solar Cells

Finally, we return to the question of what mobile ions mean for PSC efficiency more generally. To investigate this issue, we simulated JV curves as described above, but altered the perovskite layer parameters to match those described in ref. [2], in which devices achieved a certified efficiency of over 25% (see note under **Table S2**). Using these JVs, we calculated PCEs as a function of $v_S$ and $\Delta E_{TL}$ for both transport layer parameter sets, as is shown in **Figures 4a**. We find that the mobile ionic modulation of $V_{OC}$ shifts the regions of parameter space where PSCs can obtain high PCEs. Specifically, **Figure 4a** shows that the presence of mobile ions reduces the dependence of PCE on $\Delta E_{TL}$. This is because the redistribution of ions in these devices suppresses surface recombination currents and allows them to maintain a high $V_{OC}$, even at large values of $\Delta E_{TL}$ (see **Figure S15**). Thus, although the maximum achievable PCE is similar across all four simulations, devices containing mobile ions outperform those without at higher values of $\Delta E_{TL}$ (see **Figures 4b-c**). However, this improved tolerance to energetic offsets comes at the cost of a greater sensitivity to the rate of surface recombination, with devices containing mobile ions losing PCE more rapidly than those without as this parameter increases. This sensitivity arises from the constraint that mobile ions can only benefit $V_{OC}$ if $V_{OC}$ exceeds $V_{flat}$, a condition which is less likely to be met as the rate of surface recombination increases. Therefore, our results imply that, in solar cells which contain a mobile ionic species, greater gains in PCE can be made by passivating active layer/transport layer interfaces than by improving the energetic alignment between these layers. Based on the insights from the validated simulations presented here and previous results, we believe that this is one of several ways in which the design criteria for solar cells with and without a mobile ionic species differ. We summarise these differences in **Table 2**, which describes how the presence of mobile ionic charge influences the performance of a perovskite solar cell as different cell design parameters are changed relative to an equivalent device without mobile ions.

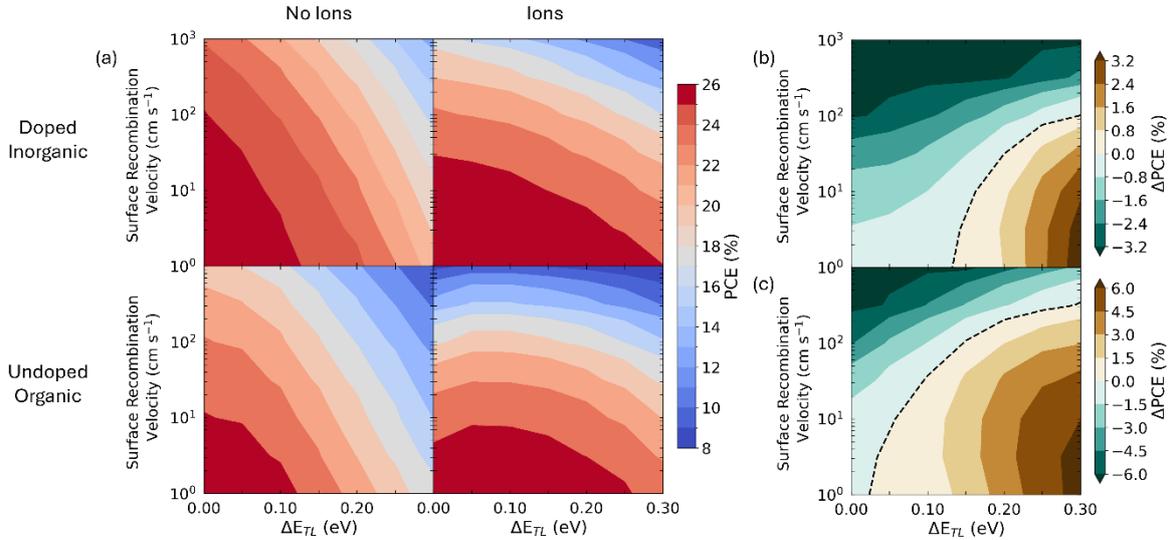

**Figure 4:** (a) A comparison of the dependence of PCE on surface recombination velocity and transport layer energetic offset ($\Delta E_{TL}$) for simulated devices without (left-hand column) and with (right-hand column) a mobile ion density of $10^{18}$ cm$^{-3}$. In the top row, we show results for devices using the doped inorganic parameter set and, in the bottom row, we show results for devices using the undoped organic parameter set. Figures (b-c) show the difference in PCE, defined as $\Delta PCE = PCE$(with ions) – PCE(no ions), for (b) the doped inorganic parameter set and (c) the undoped organic parameter set. The black dashed line is a guide for the eye to indicate where $\Delta PCE = 0$.

| Change in cell design parameter | Example | Relative impact of mobile ions on photovoltaic performance | |
|---|---|---|---|
| | | **With Ions** | **Without Mobile Ions** |
| **Reduce electron/hole diffusion length or increase perovskite thickness** | Prolonged exposure of device to light[53] | **Loss in $J_{sc}$**<br>Ions screen the built in potential, which leads to charge collection being driven by diffusion (**Figure 1a**).[11,12,54] | **Loss in fill factor**<br>Transport is drift-driven, charge collection efficiency is dependent on the applied field. (**Figure S16**). |
| **Reduce built-in potential (interface dominated recombination with low to medium rate)** | Use gold as the metallic electrode in place of silver (for same device stack) | **Gain in $V_{OC}$**<br>Ionic charge substituted for minority electronic charge carriers at key interfaces. This results in lower recombination rates when $V > V_{flat}$ (**Figures 1b-c**). | **Reduction in $V_{OC}$**<br>Minority electronic charge carriers accumulate at key interfaces. This results in increases in recombination rates when $V \leq V_{flat}$. (**Figures 1b-c**). |
| **Reduce built-in potential (interface dominated recombination with high rate)** | | **Reduction in $V_{OC}$**<br>Ionic charge attracts minority electronic carriers to key interfaces. This increases recombination rates when $V < V_{flat}$ (**Figure S3**). | **Gain in $V_{OC}$**<br>The internal electric field repels minority electronic carriers from key interfaces. This results in lower recombination rates when $V > V_{flat}$ (**Figure S3**). |
| **Reduce built-in potential (bulk dominated recombination, high injection limit, n ≈ p)** | | **Reduction in $J_{sc}$ and $V_{OC}$**<br>Ionic field screening results in greater overlap of the electron and hole populations, increasing the rate of bulk recombination processes (**Figure S3**). | **Gain in $J_{sc}$ and $V_{OC}$**<br>Presence of an electric field in the active layer reduces overlap of the electron and hole populations, decreasing the rate of bulk recombination processes (**Figure S3**) |
| **Increase energy offset between perovskite and transport layer band edges** | Replace PCBM electron transport layer with $C_{60}$ | Effect similar to reducing the built-in potential, refer to the relevant row above. | |
| | | For undoped organic transport layers, cell efficiency increases for moderate offsets due to improved extraction and smaller $V_{OC}$ losses (**Figure S17**). | For undoped organic transport layers, negligible improvement in efficiency for moderate offsets due to efficient extraction and larger $V_{OC}$ losses (**Figure S17**). |
| **Dope perovskite layer** | Vary perovskite precursor ratio[55] | All factors described above show reduced sensitivity to mobile ions due to the greater contribution of electronic charge to the perovskite layer electrostatics.[19] | |
| **Asymmetry in contact layer energetic offsets** | Use e.g., PTAA with ICBA as the charge transport layers | Asymmetric redistribution of ionic charge and injection of electronic charge from contacts results in effective doping of the perovskite bulk by uncompensated mobile ions. All factors described above show less sensitivity to mobile ions (**Supplementary Note One**).[19,22] | |

**Table 2:** Summary of the different device design parameters comparing equivalent devices with and without a mobile ionic species.

## 3. Discussion

We have shown that the presence of mobile ions does not *a priori* reduce the maximum achievable efficiency of PSCs but does change the design regime where this maximum efficiency is attained. Using drift-diffusion simulations, we demonstrate that this is because the size of interfacial recombination currents in PSCs are less sensitive to energetic misalignments at perovskite/transport layer interfaces than in equivalent devices with no mobile ions, provided that $V_{OC}$ exceeds the flat band condition in the perovskite layer. This increases the range of energetic offsets for which PSCs can maintain high PCEs, although it makes them less tolerant to high values of the surface recombination velocity, which explains the field's focus on minimising interfacial recombination in recent years. Additionally, this is the first experimental demonstration that the presence of mobile ions systematically increases steady-state $V_{OC}$ relative to devices without mobile ions. This result was found to hold for devices using n-i-p and p-i-n architectures, with MAPI or alloyed perovskite compositions, and using a range of possible transport layer materials. Most importantly, we have shown that the effects of mobile ions are still relevant in determining the steady-state performance of highly efficient PSCs, even those which show negligible hysteresis. The close agreement of our experimental and simulated results confirms the validity of drift-diffusion simulations in describing the device physics of PSCs and provides strong evidence for our claim that mobile ions are not necessarily detrimental to the performance of PSCs.

## 4. Acknowledgements

For this work, L.J.F.H., P.R.F.B. and J.R.D. thank the EPSRC ATIP project (EP/TO28513/1). F.J.A and P.D thank the EPSRC and the University of Glasgow. A.B.D, Y.L. and A.K. acknowledge support from Seed Funding for Basic Research of the University of Hong Kong.

## 5. Author Contributions

L.J.F.H and P.R.F.B. conceived the ideas. L.J.F.H. performed the drift-diffusion and interpreted simulations under the supervision of J.R.D. and P.R.F.B. Stabilise and Pulse measurements were performed by F.J.A. under the supervision of P.D. Triple cation perovskite solar cells were fabricated by Y.L and A.K under the supervision of A.B.D. L.J.F.H wrote the manuscript. All authors discussed the results and reviewed the manuscript.

## Methods

### 1. Drift-Diffusion Simulations

Device simulations were performed using Driftfusion, a software package designed to model one-dimensional, ordered semi-conductor devices which contain up to two species of mobile ions.[56] Driftfusion models ionic carriers in the perovskite layer as Schottky defects, following the results of quantum mechanical calculations by Walsh et al.[44] This means that each mobile ion is compensated for by a counter-ion of opposite charge, which is assumed to be static. Here, we included a single, positively charged mobile ionic species to reflect experimental and theoretical work which suggests that halide vacancies are the ionic carrier with the highest concentration and conductivity in metal halide perovskites and whose motion can adequately explain many aspects of device physics such as current-voltage hysteresis, inverted photocurrent transients and short-circuit electronic carrier accumulation. Additionally, we assumed that the mobile cations are restricted to the perovskite layer and neglected the electrochemical processes which may occur at perovskite/TL interfaces. Both these processes could alter interfacial recombination kinetics and/or lead to irreversible degradation of the perovskite crystal structure. The effect of such degradation is not explicitly included in our model though, assuming its consequence would be to increase surface recombination velocity, **Figure S3** suggests that this would decrease $V_{OC}$, potentially to the extent that $V_{OC} < V_{flat}$, with the result that ions start to reduce $V_{OC}$.

In order to determine an accurate value for $V_{OC}$ in the simulations which included mobile ions, it was necessary to ensure that the ions were at steady state at each point along the JV curve, which we did by using a scan rate of 0.1 mV s$^{-1}$. Thus, the JVs simulated herein are not representative of device performance as conventionally measured during a JV scan (typically done at scan rates ~10-100 mV s$^{-1}$), but instead accurately describe how the simulated device behaves once the ionic charge has re-equilibrated in response to the applied bias. This means that the $V_{OC}$ values we report are those of the fully stabilised device and any enhancement of $V_{OC}$ in the presence of mobile ions is due to their impact upon the PSC's electrostatics, as opposed to being a transient effect. JV scans on the 1.6 eV bandgap perovskite were performed at 1 Sun equivalent illumination and those on the 1.5 eV bandgap perovskite were done at 1.2 Sun equivalent, to account for the increased absorption of the narrower bandgap active layer. This was done as the $n$ and $k$ values for of the perovskite composition used in ref. [2] could not be found in the literature. Given the 3 μs Shockley-Read-Hall lifetime used in these simulations (see **Table S2**) the effective diffusion length far exceeds the layer thickness. Consequently, the differences in generation rate profile will have negligible impact on the results.

To perform the simulations in which we varied $\Delta E_{TL}$, the offsets at both interfaces were changed in a symmetric manner, while ensuring that the intrinsic carrier densities in both types of transport layer remained constant (asymmetric values of $\Delta E_{TL}$ are considered in **Supplementary Note One**). Where possible, $V_{BI}$ was set to 1.1 V, though this had to be reduced for larger $\Delta E_{TL}$ values as Driftfusion cannot model situations where the electrode work function lies within the conduction or valance band of the transport layers.

To simulate the Stabilise and Pulse protocol, we first generated a series of solutions in which we allowed the ion distribution to reach its equilibrium position for the chosen $V_{bias}$ value. Following this stabilisation, we then set the ionic mobility of each solution to zero to ensure that the ions were frozen in place during subsequent simulation. These solutions were then analysed using two different methods. In the first method, we performed an explicit simulation of the experimental protocol in which we took the input solution and ramped it to the desired $V_{pulse}$ value over a time of 0.8 ms. We then held the device at $V_{pulse}$ for 1 ms before recording the current output of the device. The ramp and dwell time were taken from the values stated in ref. [37] and this simulation protocol is akin to the one described in that work. However, this method was computationally costly, which limited the number of $V_{pulse}$ values which it was feasible to simulate, leading to a relatively low voltage resolution. Thus, we also implemented a second method in which we took the stabilised, 'frozen-ion' solutions at each $V_{bias}$ value and performed regular JV sweeps on them, the results of which agreed with those of the more explicit simulation protocol described above, and which allowed us to obtain a much higher resolution along the voltage axis (see **Figure S18**). The device efficiency for both the frozen ion and steady state JV curves was evaluated using PCE = $J_m V_m/P$ here $J_m$ and $V_m$ are the current density and voltage at the maximum power point on the JV curve, and $P$ is the light intensity incident on the device.

## 2. Experimental

### 2.1 Device Fabrication

#### 2.1.1 Triple Cation Devices (p-i-n structure)

These were fabricated as described in ref. [52]

#### 2.1.2 MAPbI$_3$ Devices (n-i-p structure)

**Materials**

All materials used were purchased from Sigma-Aldrich and used as received unless otherwise stated. Methylammonium iodide (MAI) was purchased from Greatcell Solar Ltd. 2,2′,7,7′-Tetrakis [N,N-di(4-methoxyphenyl)amino]-9,9′-spirobifluorene (spiro-OMeTAD) was purchased from Luminescence Technology Corp. Glass substrates with a conducting layer of fluorine-doped tin oxide (FTO) of 8 Ω sq$^{-1}$ sheet resistance were purchased from Yingkou Shangneng Photoelectric Material Co., Ltd.

**Device Substrate Preparation**

The FTO glass substrates were pre-patterned by laser and then cleaned sequentially using Hellmanex III, Deionised water (DI), Acetone, Ethanol and DI water, followed by a UV-Ozone plasma treatment for 15 minutes.

The compact $TiO_2$ layer was prepared by a sol-gel approach with a solution containing 0.23 M titanium isopropoxide and 0.013 M hydrochloric acid in isopropanol (IPA). 220 µL of the solution was spin-coated dynamically on top of the substrate at 2000 rpm for 45 s, dried at 150 °C for 10 min and annealed at 500 °C for 45 min.

The $SnO_2$ layer was prepared in a 1:1 solution with deionised water, stirred, and filtered through a hydrophilic PTFE filter. 50 µL of the prepared solution was dynamically spin-coated at 4000 rpm for 30s followed by annealing at 150 °C for 30 mins.

Devices with the $C_{60}$-BA ((4-(1',5'-dihydro-1'methyl-2'H-[5,6]fullereno-C60-Ih-[1,9-c]pyrrol-2'-yl)benzoic acid) interface used a 0.5 mg $mL^{-1}$ solution in chlorobenzene (CB). This solution was stirred and 50 µL was dynamically spin-coated followed by an annealing step at 100 °C for 5 min. Following the SAM layer a 0.2 wt.% IPA solution of Aluminium oxide ($Al_2O_3$) nanoparticles (Sigma-Aldrich, < 50 nm particle size, 20 wt.% in IPA) was deposited on top of the $C_{60}$-BA. Utilising 50 µL and spincoating at 2000 rpm for 30 s, samples were then dried at 120 °C for 5 min for improved nucleation of the perovskite on the SAM layer.

Following this, all substrates were immediately transferred to a nitrogen-filled glovebox.

**Device Perovskite Layer Deposition**

The perovskite solution methyl ammonium lead iodide (MAPI) was synthesised using a 1:1 solution of methyl ammonium iodide (MAI) and lead iodide ($PbI_2$) in a 4:1 dimethylformamide (DMF): dimethylsulfoxide (DMSO) solution and allowed to dissolve on a hot plate before use. The perovskite was deposited by dynamic spin coating at 1000 rpm for 10 s and then 5000 rpm for 30 s. With 50 µL of the filtered MAPI solution being deposited at the 5 s mark, followed by 300 µL of filtered chlorobenzene antisolvent. After deposition, the samples were allowed to dry on a clean cloth for 15 minutes followed by annealing at 100 °C for 15 minutes.

**Solar cell Finalisation**

For the hole transporting layer, 50 µL of a Spiro-OMeTAD solution was spin-coated on the perovskite layer at 4000 rpm for 30 s with an acceleration of 2000 rpm. Spiro-OMeTAD (90 mg/mL in chlorobenzene) was doped with 23 µL of Bis(trifluoromethane)sulfonimide lithium salt (LiTFSI) stock solution (520 mg in 1 mL acetonitrile (ACN)), 5 µL of FK-209 (Cobalt (III) salt) stock solution (180 mg in 1 mL ACN) and 35.5 µL of 4-tert-Butylpyridine (tBP) prior to spin coating. Finally, a 40 nm thick gold (Au) electrode was thermally deposited under vacuum to complete the devices. The deposition rate was 0.1 nm s$^{-1}$. Devices used in the Stabilise and Pulse rig were sealed using an epoxy adhesive mixture and a glass slide before measuring.

## 2.2 Solar Cell Characterisation

All solar cell characterisation was conducted at room temperature under ambient conditions.

### Triple Cation Devices (p-i-n structure):

JV scans were measured by a Keithley 2400 source measure unit coordinated with computer and self-made LabView program. The measurement was conducted under 1 sun illumination and AM1.5G spectrum generated by ABET Sun 2000 solar simulator through a shadow mask to define the active area (0.075 cm$^2$). The illumination intensity was calibrated by Enli PVM silicon standard reference cell. The reverse scan was from 1.2V to -0.2V while the forward scan as from -0.2V to 1.2V. The change of scan speed was achieved by changing scan step and delay time between data points. Measurement details for different JV scanning speeds are shown in the following table:

|        | Scan Step (V) | Delay Time (ms) |
|--------|---------------|-----------------|
| Normal | 0.03          | 10              |
| Fast   | 0.05          | 0.1             |
| Slow   | 0.005         | 500             |

### MAPbI$_3$ Devices (n-i-p structure):

The JV characteristics were carried out under one sun (AM 1.5) illumination using a Wavelabs Sinus-70 AAA solar simulator and measured using a Ossila Source Meter Unit. The devices were pre-biased at 1.3 V for 10 s with 1 sun illumination and measured in the reverse then forward scan direction in 0.2 V s$^{-1}$ steps. Non-reflective metal masks with an aperture area of 0.1 cm$^2$ were used to define the illumination area of the devices.

## 2.3 Stabilise and Pulse Measurements

Stabilise and Pulse measurements utilised a home-built setup and were performed on a Ossila Source Meter Unit. The source delay was set to 1 ms. A Cree High Power white LED was used as the light source. Each device was measured on the solar simulator as described in **Section 2.2** and then immediately transferred to the Stabilise and Pulse rig. We then calibrated the intensity of the white light LED to match the measured short circuit current density. When changing the bias voltage, 50 mV step increments were used with the stabilisation period being a minimum of 120 s, though in some cases this was extended to obtain a stable current output. We note that devices using the alloyed perovskite required a longer stabilisation period than those using MAPI. Consequently, the measurements were performed at ~0.5 Suns to prevent degradation during the significantly longer Stabilise and Pulse protocol.

In the main text, we show results extending up to the smallest $V_{bias}$ value necessary to determine the quasi-steady state (QSS) $V_{OC}$ as we found that the current did not stabilise prior to performing the pulsed JVs at higher values of $V_{bias}$, in contrast to what was observed for lower values of $V_{bias}$ (see **Figure S5**). Based on the fact that this effect was smaller in the devices which included $C_{60}$-BA, we believe that this instability is related to electrochemical processes which occur at the $TiO_2$/MAPI interface following prolonged illumination (though we note that the $SnO_2$/MAPI interface is similarly affected, see **Supplementary Note Four**). This interpretation is supported by our observation of a slight reduction in device performance after collecting the Stabilise and Pulse data (see **Figure S19**). However, during the pulsed JVs themselves, the ionic configuration remained stable, as we demonstrate in **Figures 2a-b** which include Stabilise and Pulse data from the forward and reverse scan directions. The negligible amount of hysteresis strongly suggests that there is no ionic motion during the pulsed JV measurement.

A $V_{flat}$ value was extracted from the Stabilise and Pulse data by analysing the gradient around the open circuit voltage of the JV obtained at each pre-bias voltage ($V_{bias}$). To find the gradient of each JV, a polynomial to the third degree was fitted to the 2-3 points above and below open circuit voltage and the gradient extracted from this fitting. Following this, we plotted the gradient around $V_{OC}$ ($dJ/dV|_{V=V_{OC}}$) against $V_{bias}$. To extract the $V_{flat}$ value from this curve, we aim to identify the inflection point. The reasoning for this is that the inflection point will be the point where a change in states is observed, or rather a flipping in device behaviour, i.e., from being below $V_{flat}$ to above $V_{flat}$. To find the inflection point, we first calculate the mean value of the steepest section of the plot by taking the average value of the two plateaus in $dJ/dV|_{V=V_{OC}}$ (one at low bias, the other at high bias, see **Figure S8**). We then perform a linear regression on the section of the $dJ/dV|_{V=V_{OC}}$ around this mean value. Finally, we extend this linear regression to find the points where it intercepts the values of $dJ/dV|_{V=V_{OC}}$ at the high and low voltage plateaus. The mean value of these intercepts is then used to estimate $V_{flat}$ for the device. We note that the reduction in device performance observed at high light intensities for the $TiO_2$ and $SnO_2$ devices without $C_{60}$-BA may lead to an underestimation of $V_{flat}$, by reducing the voltage at which the high voltage plateau in $dJ/dV|_{V=V_{OC}}$ occurs. However, we do not believe that this has significantly impacted our results as we find similar values of $V_{flat}$ in the devices which included $C_{60}$-BA, which did not show a significant reduction in performance at high forward bias. However, for this reason we have concentrated on the role of $C_{60}$-BA as a trap state passivator rather than any modulation of $V_{flat}$ in the simulations shown in **Supplementary Note Three**.

# Supplementary Information for, "More is Different: Mobile Ions Improve the Design Tolerances of Perovskite Solar Cells"

Hart, L. J. F.[*], Angus, F.J., Li, Y., Khaleed, A., Durrant, J. R., Djurišić, A. B., Docampo, P.[*], and Barnes, P. R. F.[*]

L. J. F. Hart, Dr P. R. F Barnes

Department of Physics and Centre for Processable Electronics, Imperial College London, South Kensington, U.K.

L. J. F. Hart, Professor J. R. Durrant

Department of Chemistry and Centre for Processable Electronics, Imperial College London, 82 Wood Lane, London, U.K.

F. J. Angus, Dr P. Docampo

Department of Chemistry, University of Glasgow, University Avenue, Glasgow, U.K.

Professor A. B. Djurišić, Y. Li, A. Khaleed

Department of Physics, The University of Hong Kong, Pokfulam Road, Hong Kong S.A.R, China

Professor J. R. Durrant

SPECIFIC IKC, College of Engineering, Swansea University, Bay Campus, Fabian Way, Swansea, U.K.

Email: lucy.hart18@imperial.ac.uk; pablo.docampo@glasgow.ac.uk; piers.barnes@imperial.ac.uk

**Supplementary Figures**

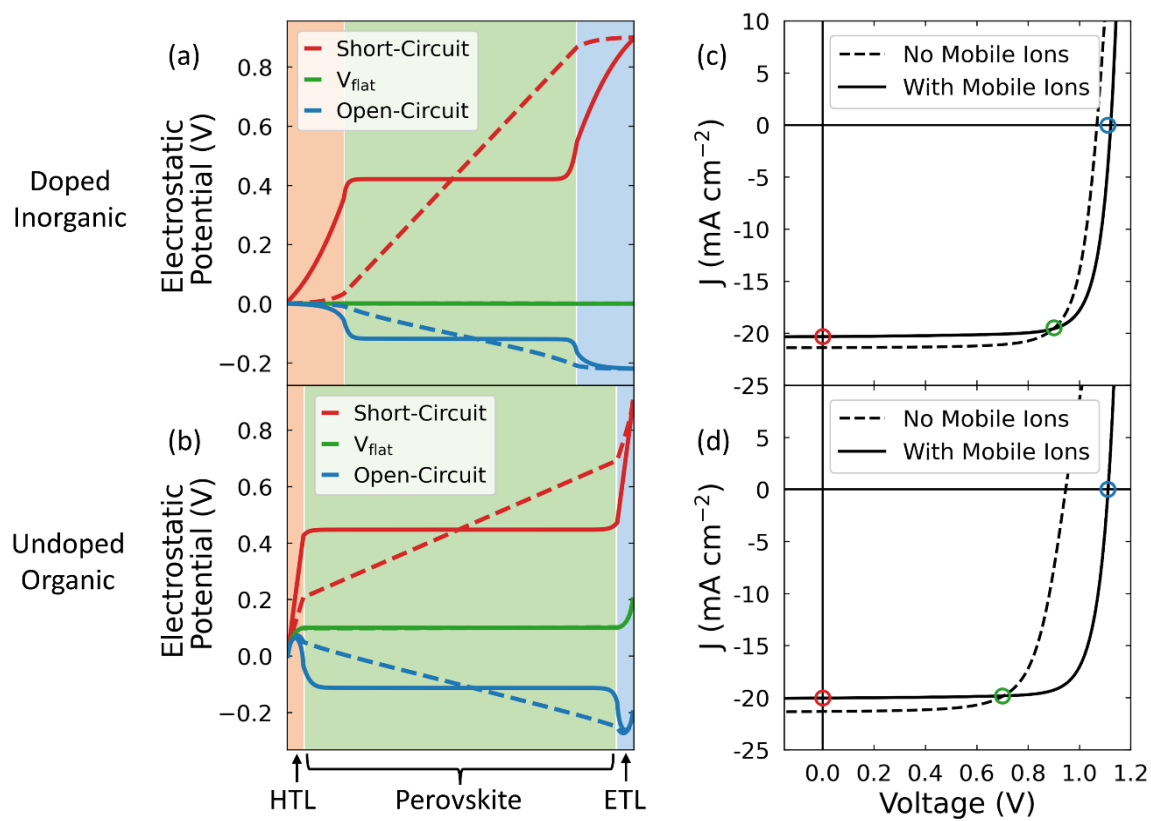

**Figure S1:** Figures (a) and (b) show the electrostatic potentials from the simulations shown in (c) and (d), which correspond to the doped inorganic and undoped organic parameters sets, respectively. Solid lines show the results when mobile ions are included ($N_{ion} = 10^{18}$ cm$^{-3}$) and dashed lines indicate results with no mobile ions. For both parameter sets, we used $\Delta E_{TL} = 0.25$ eV, $V_{BI} = 0.90$ V, and a surface recombination velocity of 10 cm s$^{-1}$. Electrostatic potentials are shown at short circuit (red, $V_{app} = 0$ V), the flat band condition in the perovskite (green, $V_{app} = 0.90$ V in figure (a) and $V_{app} = 0.70$ V in figure (b)) and open circuit for the device including mobile ions (blue, $V_{app} = 1.12$ V in figure (a) and $V_{app} = 1.11$ V in figure (b)). In both cases, we see that the inclusion of mobile ions increases the fraction of the electrostatic potential which is lost across the transport layers. Additionally, note how the use of undoped organic transport layers reduces $V_{flat}$ by 0.2 V. Figures (c) and (d) show simulated steady state JV scans performed with and without the inclusion of a mobile ionic species for the doped inorganic and undoped organic parameter sets, respectively. Red, blue, and green circles correspond to the same potentials as shown in figures (a) and (b). Note that $V_{flat}$ (as indicated by the green circles) is also the voltage at which the JV curves with and without mobile ions intersect as this is the only point in the JV scan where the electric field distribution, and thus electronic carrier distribution, are identical.

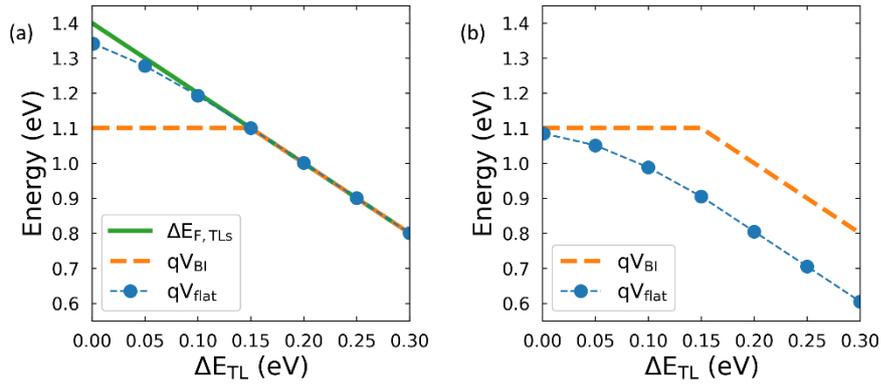

**Figure S2:** (a) The offset between the transport layers' equilibrium Fermi levels ($\Delta E_{F,TLs}$), the built in voltage ($V_{BI}$) and the flat band condition in the perovskite ($V_{flat}$) as a function of transport layers' energetic offset ($\Delta E_{TL}$) for the doped inorganic parameter set. We note that we have multiplied $V_{BI}$ and $V_{flat}$ by the elementary charge ($q$) so that they can be plotted on the same axis as $\Delta E_{F,TLs}$. This figure shows that, when the transport layers are highly conductive and have a high permittivity, $V_{flat}$ is determined by $\Delta E_{F,TLs}$. For small values of $\Delta E_{F,TLs}$, $V_{flat}$ falls slightly below $\Delta E_{F,TLs}$ and this is due to the presence of an injection barrier from the electrodes into the transport layers, which creates a depletion region at the transport layer/electrode interface. (b) $V_{BI}$ and $V_{flat}$ for the undoped organic parameter set as a function of $\Delta E_{TL}$. Since the transport layers are intrinsic, we do not show $\Delta E_{F,TLs}$ in this case. It is apparent that the use of transport layers with low conductivity and permittivity can lead to a significant reduction in $V_{flat}$ relative to $V_{BI}$, as is discussed in **Supplementary Note Two**.

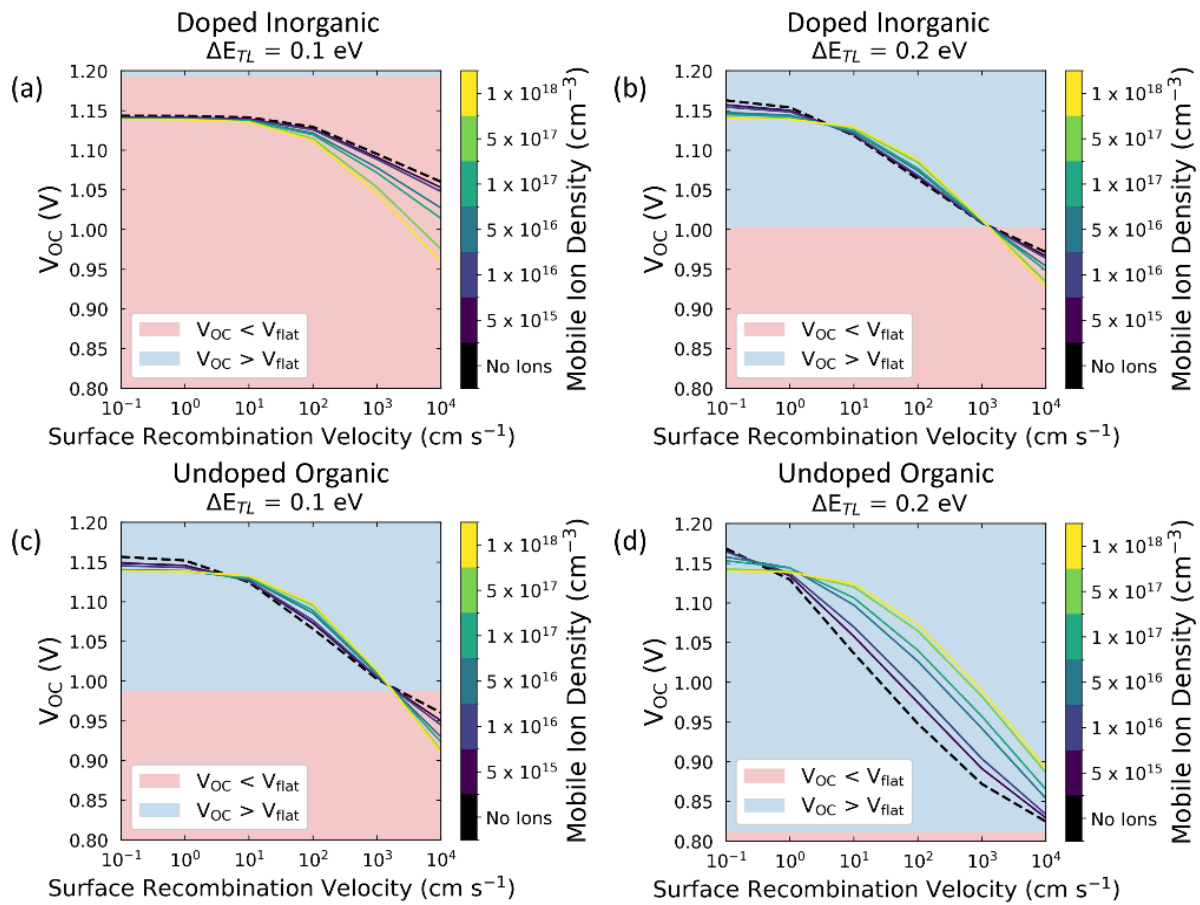

**Figure S3:** The effect of surface recombination velocity ($v_S$) on the ionic modulation of $V_{OC}$ for different combinations of parameter set and transport layer energetic offset ($\Delta E_{TL}$). Figures (a) and (b) show the case of doped inorganic transport layers for which the flat band condition across the perovskite ($V_{flat}$) is determined by the Fermi level offset between the transport layers (**Figure S2**). In (a), where there is a small $\Delta E_{TL}$, this results in $V_{OC}$ lying below $V_{flat}$ for all values of $v_S$ (in the cases of low $v_S$, the device becomes limited by Shockley-Read-Hall recombination). Under these conditions, the presence of mobile ions does not improve $V_{OC}$ as the device reaches $V_{OC}$ prior to the reversal of the electric field across the perovskite. Thus, the distribution of mobile ions always results in higher minority carrier concentrations at the perovskite/transport layer interfaces than is the case in an equivalent device without mobile ions, resulting in a higher rate of surface recombination for a given applied voltage. The situation is different in Figure (b), where $\Delta E_{TL}$ is larger thereby reducing $V_{flat}$. Under these circumstances, mobile ions can improve $V_{OC}$ via the mechanism discussed in the main text for values of $v_S$ in the range 1-1000 cms$^{-1}$. For values of $v_S \gtrsim 1000$ cms-1, we see that $V_{OC} < V_{flat}$ and so the presence of mobile ions reduces $V_{OC}$, as in Figure (a). For values of $v_S \lesssim 1$ cms$^{-1}$, we also observe that the inclusion of ions does not improve $V_{OC}$. However, in this case, it is because the recombination current becomes dominated by Shockley-Read-Hall recombination (see **Figure S4**).

Figures (c) and (d) show the case of undoped organic transport layers. This reduces $V_{flat}$ relative to the doped inorganic case (**Figure S2**) and means that it is possible to see an improvement in $V_{OC}$ due to mobile ions for a lower value of $\Delta E_{TL}$, as can be seen by contrasting Figures (a) and (c). In Figure (d), $V_{flat}$ lies below $V_{OC}$ for all values of $v_S$ due to the larger value of $\Delta E_{TL}$, meaning that the presence of mobile ions always results in an increase in $V_{OC}$ so long as surface recombination is the dominant contribution to the recombination current (which, again, is not the case for $v_S \lesssim 1$ cms$^{-1}$) . Additionally, it is interesting to note that Figures (b) and (c) show very similar trends in $V_{OC}$, despite there being a significant difference in the properties of their transport layers. We note that the choice of $\Delta E_{TL}$ for these two figures means that they have comparable values of $V_{flat}$, which demonstrates how transport layer properties play a key role in determining the distribution of the electrostatic potential in a working device. Furthermore, these figures suggests that $V_{OC}$ is largely determined by the value of $V_{flat}$ and thus that greater consideration should be paid to the capacitive properties of transport layers used in PSCs.

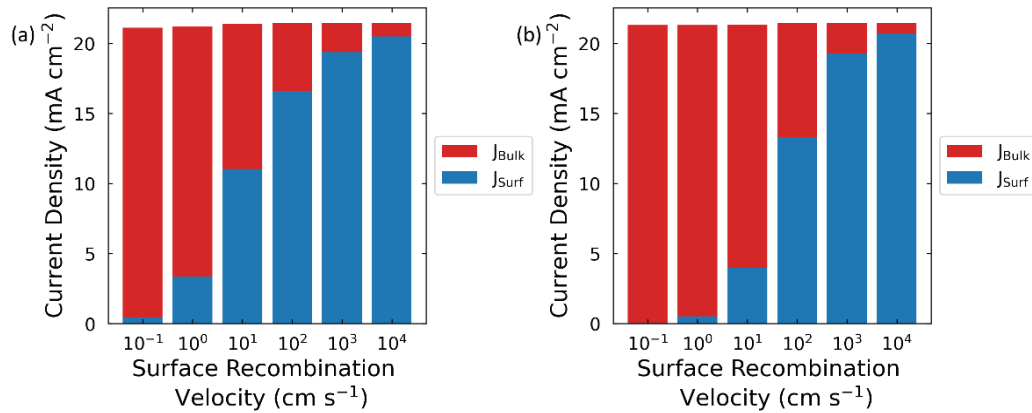

**Figure S4:** A breakdown of the non-radiative contributions to the recombination current at open circuit for the simulation results shown in **Figure S3b**. Figure (a) shows the case with no mobile ions and Figure (b) shows the case with a mobile ion concentration of $10^{18}$ cm$^{-3}$. Considering Figure (a), this device is dominated by surface recombination ($J_{Surf}$), rather than bulk recombination ($J_{Bulk}$) under open circuit conditions for values of $v_S > 10$ cm s$^{-1}$. This closely matches the point on **Figure S3b** where the devices including mobile ions start to have a higher $V_{OC}$ than those without, demonstrating that mobile ions only lead to an increase in $V_{OC}$ when surface recombination losses are dominant. We note here that, in a device with a longer bulk lifetime than that assumed in our simulations (100 ns), the presence of mobile ions would lead to an increase in $V_{OC}$ for lower values of the surface recombination velocity. Turning to Figure (b), we see that, in a device with a high mobile ion concentration, the fraction of surface recombination at open circuit is lower than in the corresponding device with no ions, up to $v_S \approx 1000$ cm s$^{-1}$. This is the point at which $V_{OC} < V_{flat}$, meaning that the presence of mobile ions acts to increase $J_{surf}$, as described in the main text.

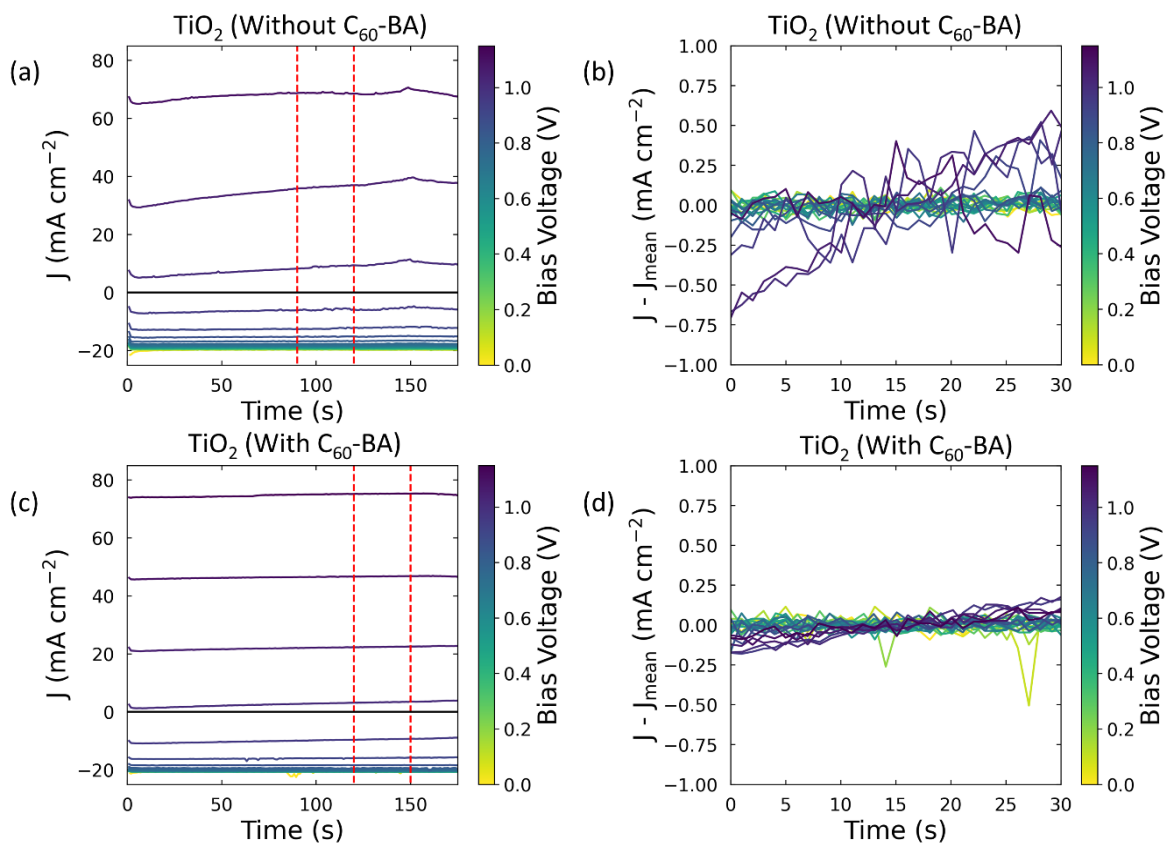

**Figure S5:** Current densities ($J$) measured during the stabilisation phase of the Stabilise and Pulse measurements. Figures (a-b) are for the devices without $C_{60}$-BA and figures (c-d) are for those with $C_{60}$-BA. In both cases, the first column shows all the data measured over the entire stabilisation period, and the vertical red lines indicate the thirty second time span over which $J$ was averaged to extract the current values for the QSS JVs. The second column focuses on these thirty seconds and shows the difference between the measured current and its mean value over this time window. It is clear that the current is less stable at higher bias voltages, though the $C_{60}$-BA layer seems to improve the device stability (see also **Figure S18**). This suggests that the observed change in $J$ at longer times is linked to processes which occur at the perovskite/TiO$_2$ interface, rather than the movement of mobile ionic charge.

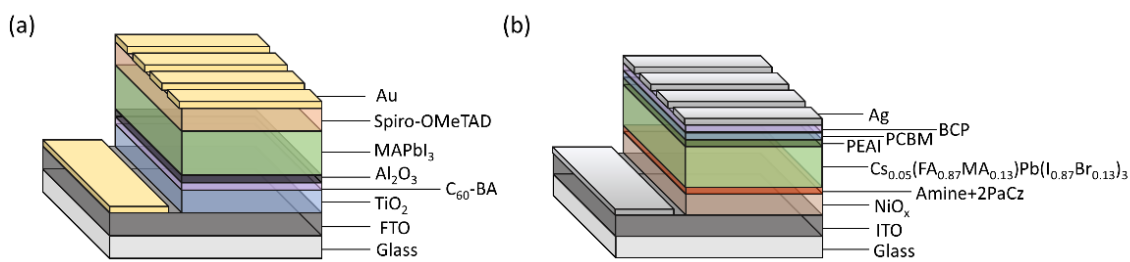

**Figure S6:** Device stacks for (a) the n-i-p MAPbI$_3$ devices used to measure the results shown in **Figure 2** of the main text and (b) the p-i-n triple cation device used to measure the results shown in **Figure 3** of the main text.

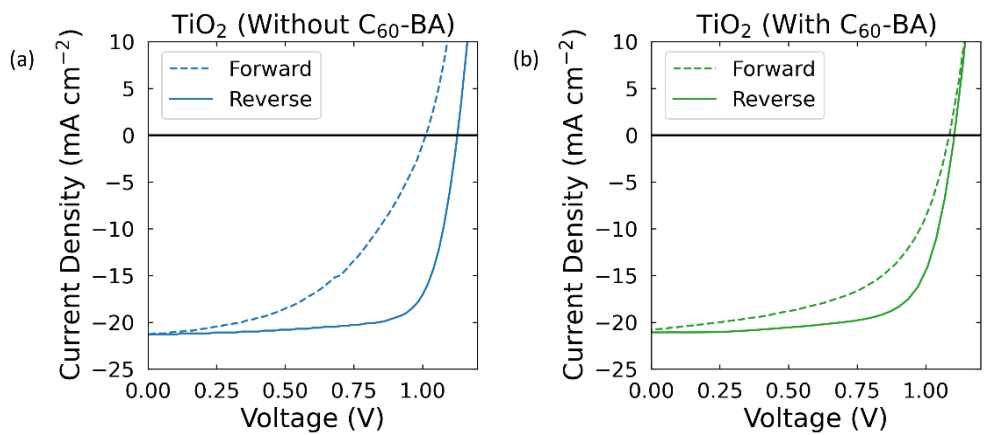

**Figure S7:** JV curves measured under the Solar Sim using the protocol described in the **Methods** for (a) the Au/Spiro-OMeTAD/MAPI/TiO$_2$/FTO device stack and (b) the Au/Spiro-OMeTAD/MAPI/C$_{60}$-BA/TiO$_2$/FTO device stack. JV parameters are given in **Table S2** and **Table S3**, respectively.

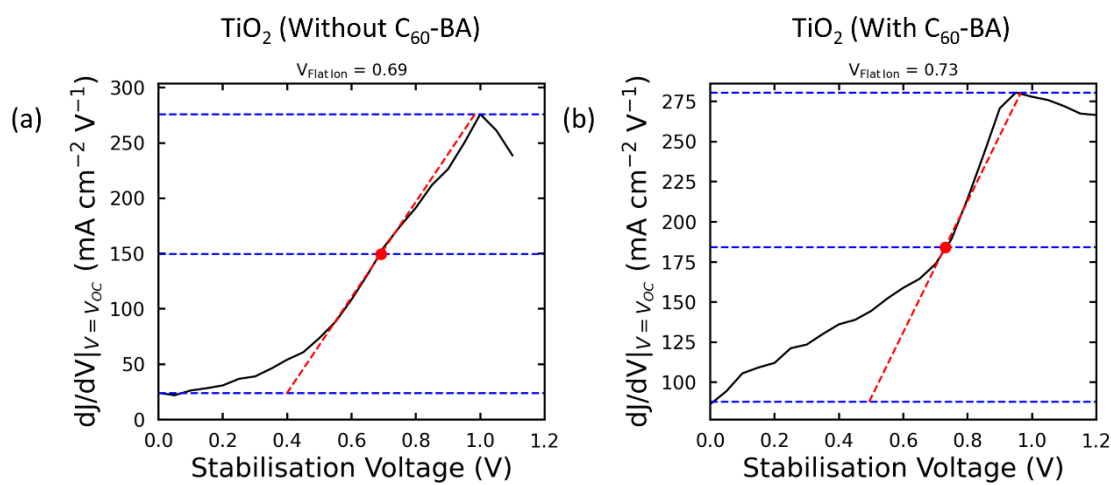

**Figure S8:** Plots used to extract the value of $V_{flat}$ from the Stabilise and Pulse data for (a) the device without C$_{60}$-BA and (b) the device with C$_{60}$-BA. Details of the fitting procedure are given in the **Methods** section.

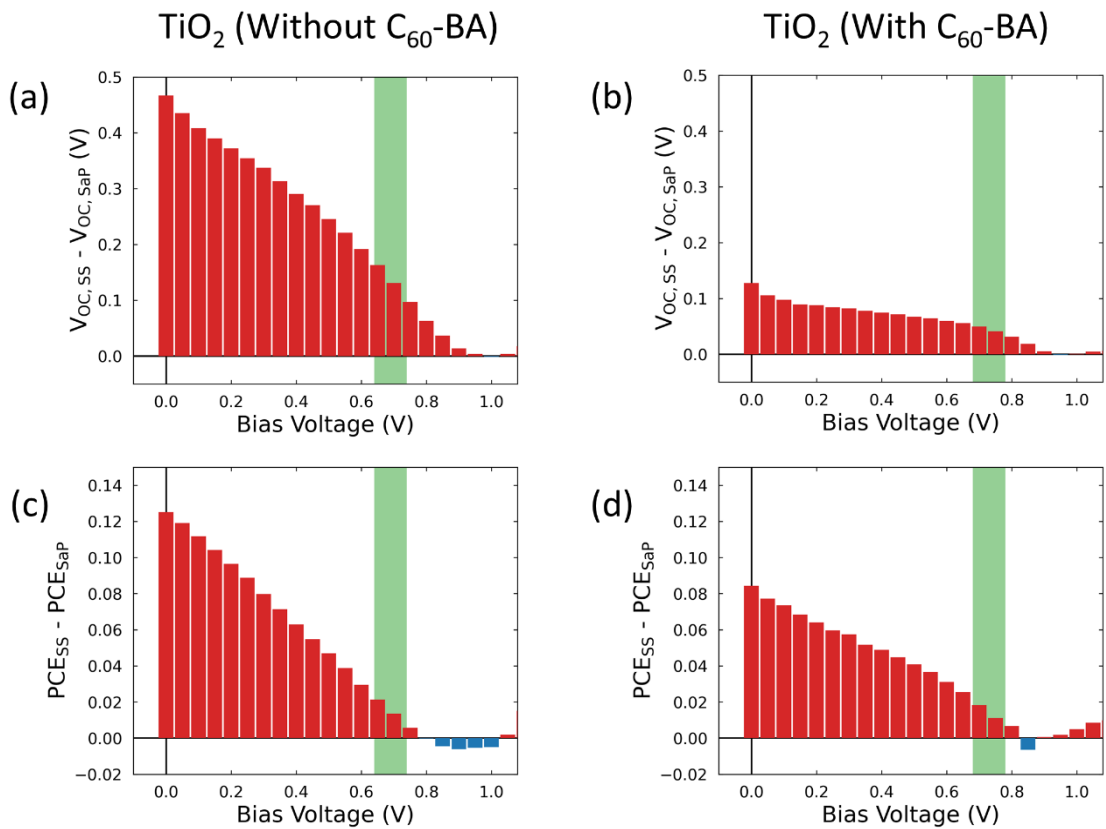

**Figure S9:** Difference between (a-b) $V_{OC}$ and (c-d) PCE as measured from the JVs with the ions at quasi-steady state and the JVs as measured using the Stabilise and Pulse (SaP) technique at different bias voltages. The left-hand column shows the device without $C_{60}$-BA and the right-hand column shows the device with $C_{60}$-BA. In all cases, the green shaded region indicates the value extracted for $V_{flat}$ (see **Methods** and **Figure S8**) with the associated error ($\pm 0.05$ V).

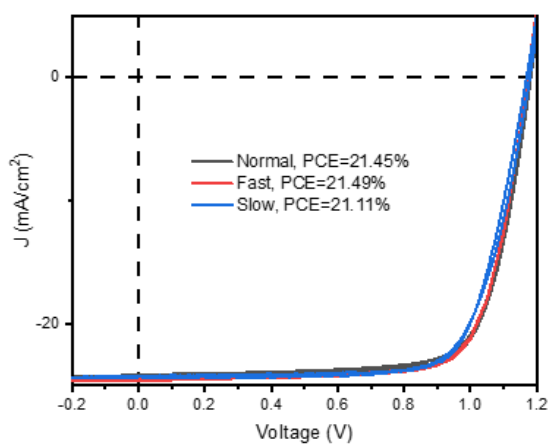

| Scan Speed | Direction | $V_{OC}$ (V) | $J_{SC}$ (mA cm$^{-2}$) | PCE (%) | FF |
|---|---|---|---|---|---|
| Normal | Reverse | 1.17 | 24.00 | 21.45 | 0.76 |
| | Forward | 1.19 | 24.00 | 21.13 | 0.73 |
| Fast | Reverse | 1.15 | 24.58 | 21.49 | 0.76 |
| | Forward | 1.15 | 24.58 | 21.59 | 0.76 |
| Slow | Reverse | 1.18 | 24.33 | 20.90 | 0.73 |
| | Forward | 1.17 | 24.33 | 21.11 | 0.74 |

**Figure S10:** JV scans of the ITO/NiO$_x$/Amine/2PaCz/perovskite/PEAI/PCBM/BCP/Ag device shown in **Figure 3** of the main text. The perovskite composition is Cs$_{0.05}$(FA$_{0.87}$MA$_{0.13}$)Pb(I$_{0.87}$Br$_{0.13}$)$_3$ and the full device stack is shown in **Figure S6**. JV parameters are provided in the table and see the **Methods** for a description of the measurement protocol used to measure the JV scans.

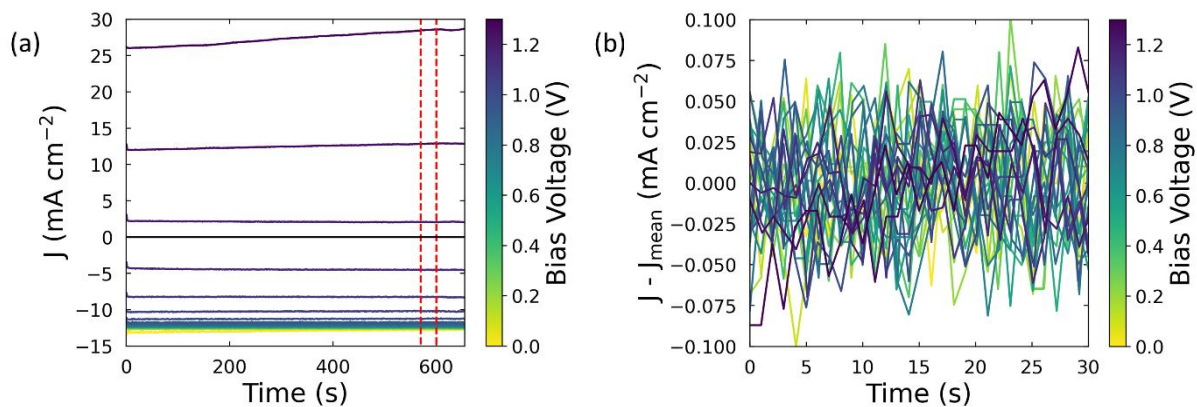

**Figure S11:** Stabilisation data for the Stabilise and Pulse measurement shown in **Figure 4** of the main text. (a) Current density ($J$) measured during the stabilisation phase of the Stabilise and Pulse measurements for all measured values of $V_{bias}$. The vertical red lines indicate the thirty second time span over which $J$ was averaged to extract the current values for the quasi-steady state (QSS) JVs. (b) The current measured in the thirty second time period indicated in (a) minus its mean value.

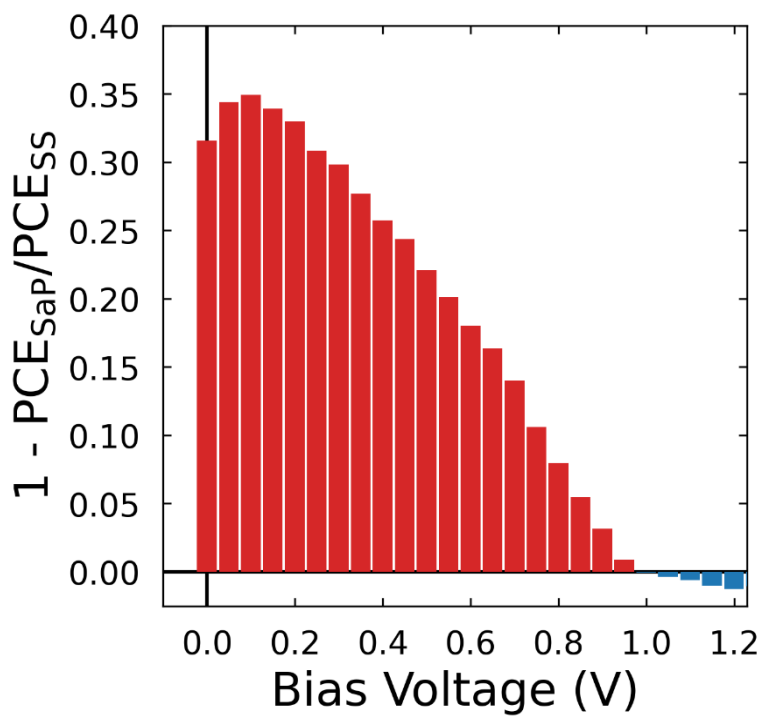

**Figure S12**: The normalised difference between the PCE of the JV curve where the ions remain at quasi-steady state (PCE$_{SS}$) and those measured using the Stabilise and Pulse protocol at different values of $V_{bias}$ (PCE$_{SaP}$). We have normalised this data as, for measurements on devices using the triple cation perovskite composition, the intensity of the LED light was calibrated such that the device gave half the $J_{SC}$ value as measured under the Solar Simulator (see main text). However, we did not explicitly measure this intensity and so cannot accurately convert our data to a PCE. Regardless, this plot indicates that mobile ions improve the device's PCE if $V_{flat} < 1.0$ V. Our simulation results indicate that $V_{flat} \approx 0.85$ eV, corresponding to a PCE improvement of ~ 1% due to mobile ions if we assume an intensity of 50 mWcm$^{-2}$.

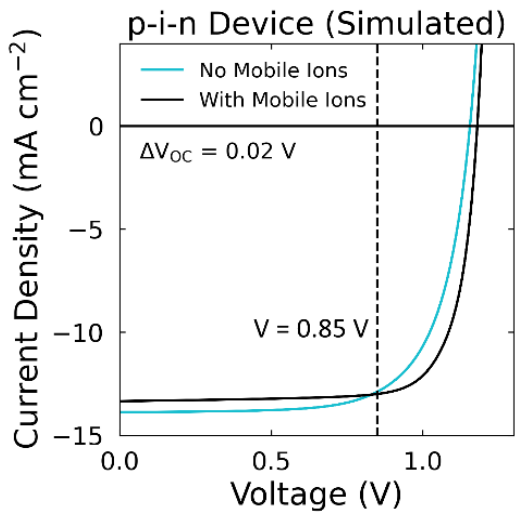

**Figure S13:** Simulated JV curves obtained using the parameters listed in **Table S5** with and without the inclusion of a mobile ionic species. The black dashed line indicates V = 0.85 V which is the voltage at which the 'No Mobile Ions' and 'With Mobile Ions' JV curves intersect. As commented upon in the caption to **Figure S1**, this intersection occurs at $V_{flat}$, thereby demonstrating that $V_{flat}$ = 0.85 V for this device.

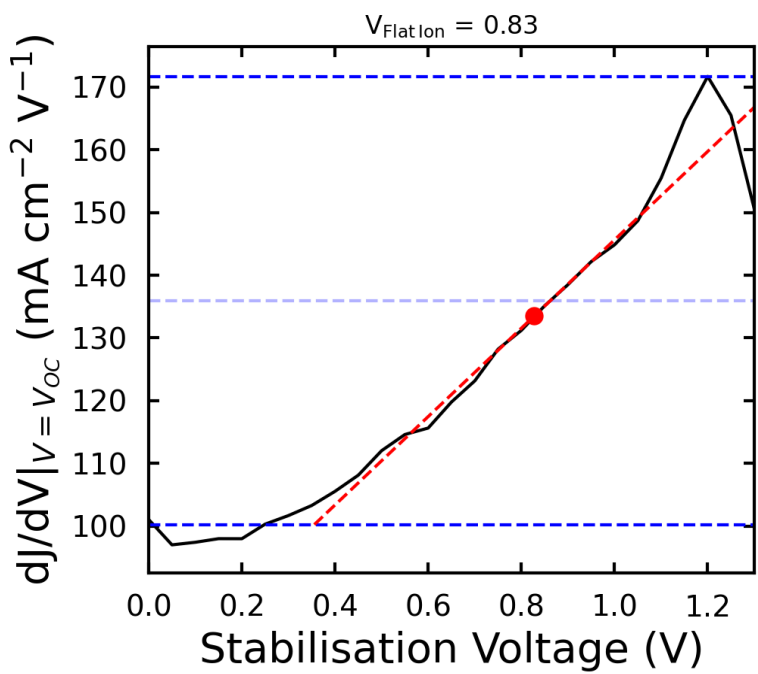

**Figure S14:** Plot used to extract the value of $V_{flat}$ from the Stabilise and Pulse data for the p-i-n device shown in **Figure 4a** of the main text. Details of the fitting procedure are given in the **Methods** section.

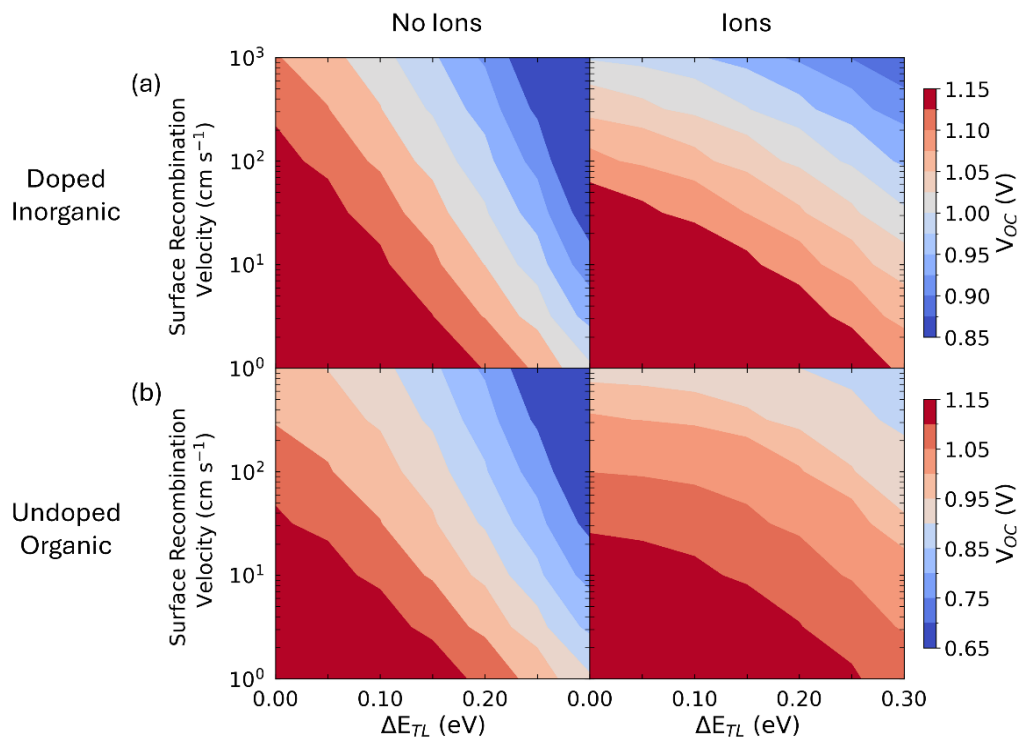

**Figure S15:** $V_{OC}$ of simulated devices shown in **Figure 4** as a function of surface recombination velocity and $\Delta E_{TL}$ for (a) the doped inorganic parameter set without (left-hand column) and with (right-hand column) the inclusion of mobile ions and (b) the undoped organic parameter set without (left-hand column) and with (right-hand column) the inclusion of mobile ions. When mobile ions were included, we used a mobile ion concentration of $10^{18}$ cm$^{-3}$. By comparing the two columns, we can see that the inclusion of mobile ions in the simulations reduces the dependence of $V_{OC}$ on $\Delta E_{TL}$, and that this effect is particularly pronounced for the undoped organic parameter set.

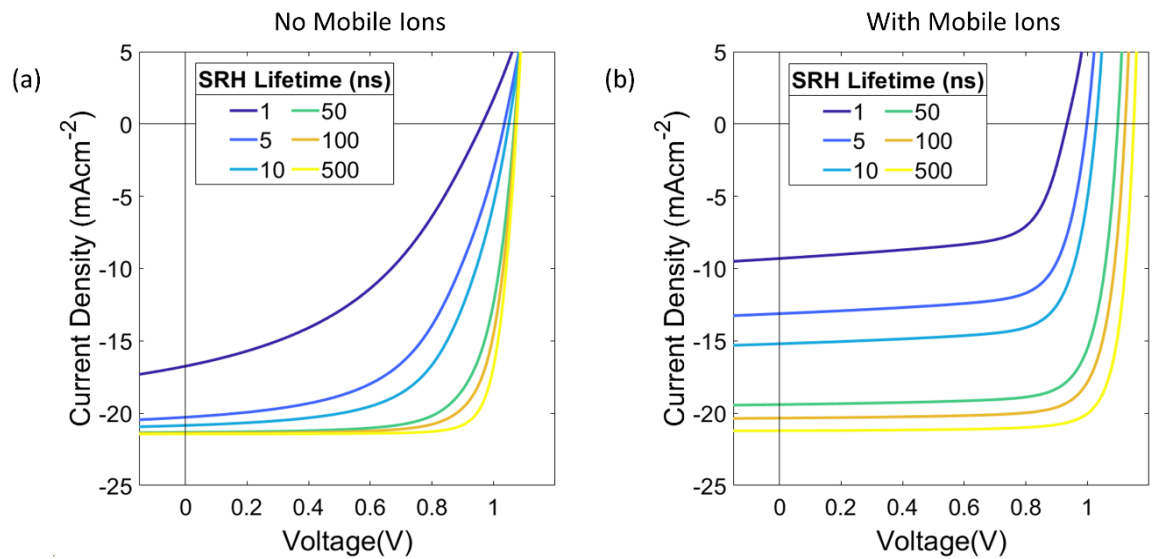

**Figure S16:** The dependence of JV curves on bulk Shockley-Read-Hall (SRH) lifetime for the doped inorganic parameter set with a fixed energetic offset of 0.25 eV and (a) no mobile ions and (b) a mobile ion density of $10^{18}$ cm$^{-3}$. The case with mobile ions was simulated such that the ions remained at quasi-steady state. The trap states were assumed to lie midgap in the perovskite and electrons and holes were given the same SRH lifetime. The presence of mobile ions increases the loss in photocurrent at low bias voltages, but this device maintains a high fill factor since field screening by the mobile ionic species reduces the dependence of extraction efficiency on the applied bias.

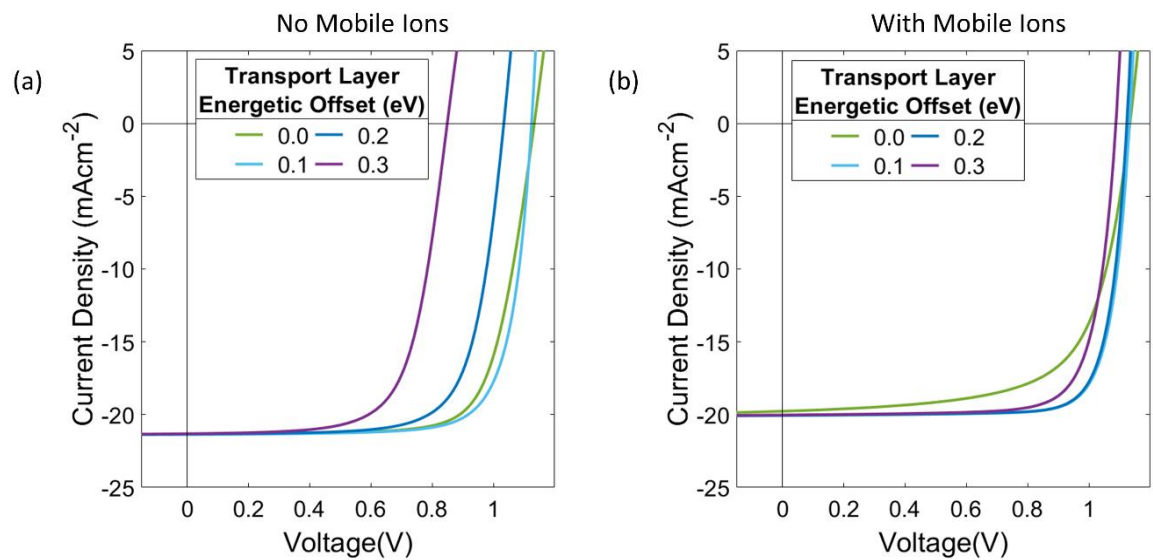

**Figure S17:** The dependence of JV curves on transport layer energetic offset (varied symmetrically) for the undoped organic parameter set with (a) no mobile ions and (b) a mobile ion density of $10^{18}$ cm$^{-3}$. The case with mobile ions was simulated such that the ions remained at quasi-steady state. In both devices, the fill factor improves upon increasing the energetic offset from 0.0 eV to 0.1 eV, but the improvement is greater in the case with mobile ions. Additionally, this figure illustrates how $V_{OC}$ is far less sensitive to transport layer energetic offsets in devices with a mobile ionic species (subject to the caveats discussed in the main text and **Figure S3**).

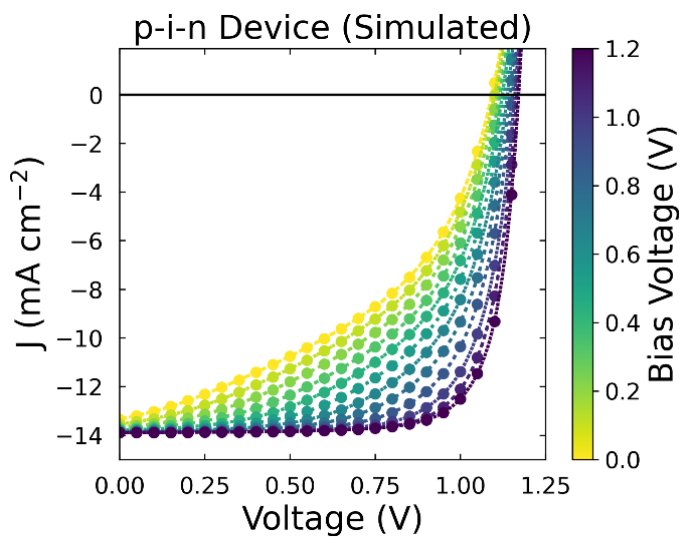

**Figure S18:** Figure to show that the two protocols which are described in the **Methods** section and used to simulate the Stabilise and Pulse measurements yield equivalent results. The solid circles show the results of the explicit simulations of the Stabilise and Pulse procedure, whereas the dashed lines show the results of JV scans performed with the ion mobility set to zero following a prebiasing period at the desired bias voltage. It can be seen that both simulation protocols yield equivalent results, but the JV scans allow for a higher voltage resolution along the *x*-axis.

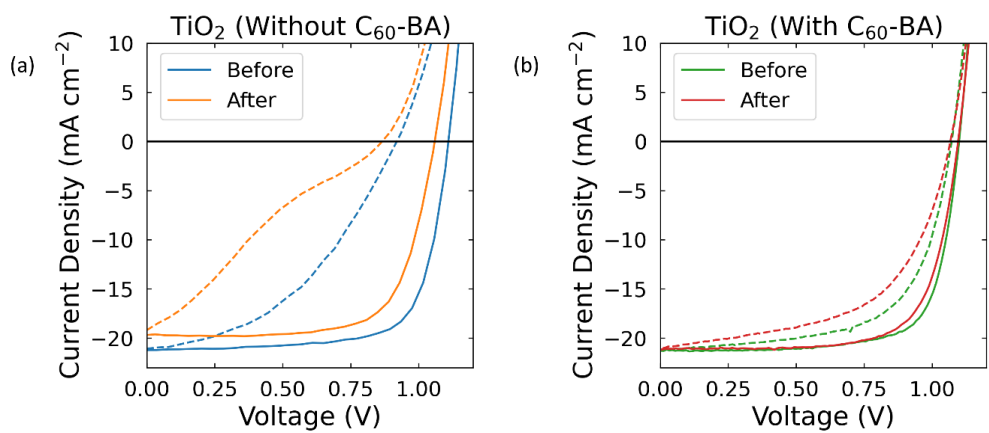

**Figure S19:** JV curves before and after the Stabilise and Pulse measurement for the devices shown in **Figure 2** of the main text. Solid lines indicate the reverse scan, and dashed lines the forward scan. We note that these JV scans were not performed under the Solar Sim, but under the LED illumination of the Stabilise and Pulse set-up, which is why the JV curves measured before the Stabilise and Pulse measurement do not perfectly match those shown in **Figure S6**. The device without the $C_{60}$-BA interlayer shows greater degradation after the measurement, which we believe is linked to the instability in the device's current which was observed at high values of $V_{bias}$ (see **Figure S4**).

## Supplementary Tables

**Table S1:** Summary of the simulation parameters for the transport layers. When not otherwise stated, the same parameters were used for the electron and hole transport layers (ETL and HTL, respectively). Recombination parameters (Shockley-Read Hall Lifetimes and radiative recombination rate) were set such that there were no recombination losses within the transport layers, except recombination at interfaces with the perovskite and electrode. $E_C$ and $E_V$ refer to the conduction and valance band energies, respectively.

| Parameter | Doped Inorganic | Undoped Organic |
|---|---|---|
| Bandgap | 2.50 eV | 2.50 eV |
| ETL Fermi Level | $E_C - 0.1$ eV | Midgap |
| HTL Fermi Level | $E_V + 0.1$ eV | Midgap |
| Energy Level of Trap States | Midgap | Midgap |
| Thickness | 100 nm | 20 nm |
| Carrier Mobility | 10 cm$^2$ V$^{-1}$ s$^{-1}$ | 0.001 cm$^2$ V$^{-1}$ s$^{-1}$ |
| Relative Permittivity | 50 | 3.5 |
| Effective Density of States | $5 \times 10^{18}$ cm$^{-3}$ | $1 \times 10^{20}$ cm$^{-3}$ |

**Table S2:** Summary of the simulation parameters for the perovskite layer which were used to generate the results shown in **Figure 2** of the main text. Monomolecular trap states were assumed to lie in the middle of the bandgap and the perovskite was treated as an intrinsic semiconductor. At the interfaces, trap states were placed midgap for the perovskite layer and thus surface recombination velocities were assumed to be equal for electrons and holes.

| Parameter | Value | Ref |
|---|---|---|
| Perovskite Bandgap | 1.60 eV | [1] |
| Perovskite Valance Band Energy | -5.4 eV | [2] |
| Perovskite Thickness | 400 nm | [a] |
| Perovskite Carrier Mobility | 1.0 cm$^2$ V$^{-1}$ s$^{-1}$ | [3] |
| Shockley-Read-Hall Lifetime | 100 ns | [4] |
| Radiative Recombination Rate | 5 x 10$^{-11}$ cm$^{-3}$ s$^{-1}$ | [5] |
| Perovskite Relative Permittivity | 25 | [6] |
| Perovskite Effective Density of States | 5 x 10$^{18}$ cm$^{-3}$ | [7] |

a) Typical thickness for perovskite layer.

To simulate the contour plots shown in **Figure 4** of the main text we changed the perovskite bandgap and Shockley-Read-Hall lifetime to match the values given in **Ref. 2** of the main text which used FAPbI$_3$ for the perovskite layer, as opposed to MAPbI$_3$. This meant that we reduced the bandgap to 1.50 eV (while maintaining an intrinsic perovskite layer with midgap trap states) and we increased the Shockley-Read-Hall lifetime to 3 μs. Additionally, as we could not obtain *n* and *k* values for FAPbI$_3$, simulations were done at 1.2 Suns to match the $J_{SC}$ values reported in **Ref. 2**. All other simulation parameters remained as reported in **Tables S1-2**.

As noted in the main text, the built in potential across the device was held at 1.1 V and we used symmetric energetic offsets between the electrode work function and the relevant band edge of the transport layers. However, for transport layer-perovskite energetic offsets > 0.15 eV, it was not possible to maintain V$_{BI}$ at 1.1 V as this would have caused the electrode's work function to lie within the conduction (valance) band of the ETL (HTL). In these cases, the work functions were reduced such that they always lay 0.1 eV below the relevant band edge. This ensured an Ohmic contact between the electrodes and the transport layers in the case of the doped inorganic parameter set.

In simulations where the surface recombination velocity was not varied, we used a default value of 10 cm s$^{-1}$, as this value allowed us to access regimes dominated by either SRH recombination or surface recombination within the range of $\Delta E_{TL}$ values investigated in this work.

**Table S3:** Summary of the JV parameters for the Au/Spiro-OMeTAD/MAPbI$_3$/TiO$_2$/FTO devices discussed in the main text. JV parameters are given for three cases: as measured under the Solar Simulator using the protocol described in the **Methods** section (parameters taken from the reverse scan), as measured with the mobile ions at quasi-steady state (QSS JV) and as measured using Stabilise and Pulse measurements carried out at $V_{bias} = V_{flat}$ (SaP JV). The parameters corresponding to the devices shown in the main text are given in bold and data for all other devices is shown in **Supplementary Note Five**. Additionally, we provide the value of $V_{flat}$ extracted from our analysis of the Stabilise and Pulse data in column 2.

| Device Number | $V_{flat}$ (V) | Method | $J_{SC}$ (mA cm$^{-2}$) | $V_{OC}$ (V) | FF | PCE (%) |
|---|---|---|---|---|---|---|
| **1** | **0.69 ± 0.05** | **Solar Sim** | **21.30** | **1.127** | **0.75** | **17.90** |
| | | **QSS JV** | **19.72** | **1.019** | **0.71** | **14.26** |
| | | **SaP JV** | **21.34** | **0.856** | **0.66** | **12.13** |
| 2 | 0.66 ± 0.05 | Solar Sim | 20.78 | 1.081 | 0.71 | 16.00 |
| | | QSS JV | 18.60 | 0.989 | 0.64 | 11.75 |
| | | SaP JV | 21.49 | 0.827 | 0.61 | 10.80 |
| 3 | 0.64 ± 0.05 | Solar Sim | 21.89 | 1.080 | 0.70 | 16.47 |
| | | QSS JV | 20.45 | 1.005 | 0.62 | 12.81 |
| | | SaP JV | 22.79 | 0.823 | 0.62 | 11.64 |
| 4 | 0.67 ± 0.05 | Solar Sim | 21.82 | 1.082 | 0.71 | 16.68 |
| | | QSS JV | 18.98 | 1.008 | 0.60 | 11.47 |
| | | SaP JV | 21.30 | 0.855 | 0.62 | 11.38 |
| 5 | 0.65 ± 0.05 | Solar Sim | 20.25 | 1.073 | 0.69 | 15.01 |
| | | QSS JV | 18.59 | 0.991 | 0.61 | 11.19 |
| | | SaP JV | 21.44 | 0.846 | 0.57 | 10.34 |

**Table S4:** Summary of the JV parameters for the Au/Spiro-OMeTAD/MAPI/C$_{60}$-BA/TiO$_2$/FTO devices discussed in the main text. The parameters corresponding to the devices shown in the main text are given in bold and data for all other devices is shown in **Supplementary Note Five**. For further details, see the caption of **Table S3**.

| Device Number | $V_{flat}$ (V) | Method | $J_{SC}$ (mA cm$^{-2}$) | $V_{OC}$ (V) | FF | PCE (%) |
|---|---|---|---|---|---|---|
| 1 | 0.76 ± 0.05 | SolarSim | 21.08 | 1.104 | 0.71 | 16.55 |
| | | QSS JV | 19.89 | 1.027 | 0.79 | 16.10 |
| | | SaP JV | 21.88 | 0.987 | 0.68 | 14.78 |
| 2 | 0.75 ± 0.05 | SolarSim | 21.24 | 1.081 | 0.73 | 16.69 |
| | | QSS JV | 20.25 | 0.997 | 0.67 | 13.48 |
| | | SaP JV | 21.41 | 0.973 | 0.63 | 13.04 |
| 3 | 0.72 ± 0.05 | SolarSim | 21.70 | 1.074 | 0.74 | 17.24 |
| | | QSS JV | 21.10 | 1.005 | 0.72 | 15.32 |
| | | SaP JV | 21.79 | 0.970 | 0.67 | 14.17 |
| 4 | 0.70 ± 0.05 | SolarSim | 20.34 | 1.078 | 0.71 | 15.67 |
| | | QSS JV | 19.85 | 1.006 | 0.74 | 14.75 |
| | | SaP JV | 20.22 | 0.980 | 0.70 | 13.95 |
| **5** | **0.73 ± 0.05** | **SolarSim** | **21.30** | **1.099** | **0.74** | **17.34** |
| | | **QSS JV** | **20.68** | **1.037** | **0.77** | **16.55** |
| | | **SaP JV** | **21.45** | **0.996** | **0.72** | **15.46** |
| 6 | 0.74 ± 0.05 | SolarSim | 21.85 | 1.068 | 0.73 | 17.09 |
| | | QSS JV | 20.73 | 0.986 | 0.64 | 13.16 |
| | | SaP JV | 22.04 | 0.937 | 0.64 | 13.18 |

**Table S5:** Parameters set used to simulate the Stabilise and Pulse measurements performed on the high-performance p-i-n device, the results of which are shown in **Figure 3c** of the main text. The perovskite was assumed to be intrinsic and trap states (both bulk and interfacial) were assumed to lie midgap. Where no comment or reference is given, parameters were tuned to qualitatively reproduce the behaviour observed in the Stabilise and Pulse measurements. We have assumed that the dominant effect of the interfacial PEAI and 2PACz layers (see device stack shown in **Figure S6**) is to reduce surface recombination, thereby motivating our use of low values for these parameters. Measurements were simulated under 0.65 suns so as to match the experimentally measured $J_{SC}$ values.

| Parameter | Value | Ref |
|---|---|---|
| Perovskite Bandgap | 1.60 eV | [8] |
| Perovskite Valance Band Energy | -5.5 eV | [8] |
| Perovskite Thickness | 400 nm | a |
| Perovskite Carrier Mobility | 0.25 cm$^2$ V$^{-1}$ s$^{-1}$ | [3] |
| Shockley-Read-Hall Lifetime | 400 ns | - |
| Radiative Recombination Rate | 5 x 10$^{-11}$ cm$^{-3}$ s$^{-1}$ | [5] |
| Perovskite Relative Permittivity | 25 | [6] |
| Perovskite Effective Density of States | 5 x 10$^{18}$ cm$^{-3}$ | [7] |
| Perovskite Mobile Ion Density | 7 x 10$^{16}$ cm$^{-3}$ | [6] |
| ETL Bandgap | 2.00 eV | [9] |
| ETL Conduction Band Energy | -4.1 eV | b |
| ETL Fermi Level | -5.1 eV | c |
| ETL Thickness | 25 nm | d |
| ETL Carrier Mobility | 0.005 cm$^2$ V$^{-1}$ s$^{-1}$ | [10] |
| ETL Relative Permittivity | 3.5 | e |
| ETL Effective Density of States | 1 x 10$^{20}$ cm$^{-3}$ | [11] |
| Surface Recombination Velocity of Holes at the Perovskite/ETL Interface | 0.65 cms$^{-1}$ | - |
| Surface Recombination Velocity of Electrons at the Perovskite/ETL Interface | 10$^7$ cms$^{-1}$ | f |
| Cathode Work Function | -4.175 V | - |
| HTL Bandgap | 3.00 eV | g |
| HTL Valance Band Energy | -5.3 eV | h |
| HTL Fermi Level | -5.1 eV | [12] |
| HTL Thickness | 80 nm | [12] |

| HTL Carrier Mobility | 0.01 cm$^2$ V$^{-1}$ s$^{-1}$ | 13 |
|---|---|---|
| HTL Relative Permittivity | 12 | 14 |
| HTL Effective Density of States | 1 x 10$^{19}$ cm$^{-3}$ | i |
| Surface Recombination Velocity at the of Electrons Perovskite/HTL Interface | 0.1 cms$^{-1}$ | - |
| Surface Recombination Velocity at the of Holes Perovskite/HTL Interface | 10$^7$ cms$^{-1}$ | - |
| Anode Work Function | -5.225 V | - |

a) Typical thickness for perovskite layer.

b) Values in the range -4.2 eV to -3.9 eV have been reported for PCBM.[9,15] We chose the value in this range which best reproduced the Stabilise and Pulse measurements.

c) PCBM was assumed to be an intrinsic semiconductor.

d) Typical thickness for PCBM ETLs.

e) Typical value for organic semiconductors.

f) Setting a high value for the majority carrier surface recombination velocity was observed to have a negligible effect on simulated JV curves with the ions at quasi-steady state. However, this parameter had a large effect on the shape of the simulated Stabilise and Pulse curves at low bias voltages. Thus, it's value was varied to match the experimental results, with larger values found to lead to less 's-shaped' JV curves at low bias voltages.

g) Bandgap of NiO$_x$ varies depending on fabrication route, but typically > 3.0 eV. Further increases in the band gap yielded no change in the simulation results, but increased the time taken per simulation and so this parameter was set to 3 eV.

h) Values in the range -5.45 eV to -5.05 eV have been reported for NiO.[12,16] We chose the value in this range which best reproduced the Stabilise and Pulse measurements.

i) Calculated using parabolic band approximation and m$^*$ = m$_e$.[17]

**Table S6:** Parameter set used to simulate the Stabilise and Pulse measurements performed on the Au/Spiro-OMeTAD/MAPbI$_3$/C$_{60}$-BA/TiO$_2$/FTO device stacks, the results of which are shown in **Supplementary Note Three**. In these simulations, the MAPbI$_3$ was assumed to be intrinsic and trap states (both bulk and interfacial) were assumed to lie midgap. Where no comment or reference is given, parameters were tuned to qualitatively reproduce the behaviour observed in the Stabilise and Pulse measurements.

| Parameter | Value | Ref |
|---|---|---|
| Perovskite Bandgap | 1.60 eV | [1] |
| Perovskite Valance Band Energy | -5.4 eV | [2] |
| Perovskite Thickness | 400 nm | a |
| Perovskite Carrier Mobility | 3 cm$^2$ V$^{-1}$ s$^{-1}$ | [18] |
| Shockley-Read-Hall Lifetime | 25 ns | - |
| Radiative Recombination Rate | 5 x 10$^{-11}$ cm$^{-3}$ s$^{-1}$ | [5] |
| Perovskite Relative Permittivity | 25 | [6] |
| Perovskite Effective Density of States | 5 x 10$^{18}$ cm$^{-3}$ | [7] |
| Perovskite Mobile Ion Density | 1 x 10$^{17}$ cm$^{-3}$ | - |
| ETL Bandgap | 3.00 eV | b |
| ETL Conduction Band Energy | -4.0 eV | c |
| ETL Fermi Level | -4.2 eV | - |
| ETL Thickness | 25 nm | [19] |
| ETL Carrier Mobility | 0.1 cm$^2$ V$^{-1}$ s$^{-1}$ | d |
| ETL Relative Permittivity | 19 | e |
| ETL Effective Density of States | 1 x 10$^{19}$ cm$^{-3}$ | f |
| Surface Recombination Velocity of Holes at the Perovskite/ETL Interface | 50 cms$^{-1}$ (without C$_{60}$-BA) 0.05 cms$^{-1}$ (with C$_{60}$-BA) | - |
| Surface Recombination Velocity of Electrons at the Perovskite/ETL Interface | 1000 cms$^{-1}$ | g |
| Cathode Work Function | -4.2 V | - |
| HTL Bandgap | 3.00 eV | h |
| HTL Valance Band Energy | -5.05 eV | i |
| HTL Fermi Level | -4.85 eV | j |
| HTL Thickness | 200 nm | [20] |
| HTL Carrier Mobility | 0.01 cm$^2$ V$^{-1}$ s$^{-1}$ | j |

| HTL Relative Permittivity | 3.5 | k |
|---|---|---|
| HTL Effective Density of States | $1 \times 10^{20}$ cm$^{-3}$ | l |
| Surface Recombination Velocity of Electrons at the Perovskite/HTL Interface | 10 cms$^{-1}$ | - |
| Surface Recombination Velocity of Holes at the Perovskite/HTL Interface | 1000 cms$^{-1}$ | g |
| Anode Work Function | -4.85 V | - |

a) Typical thickness for perovskite layer.

b) Bandgap of $TiO_2$ varies depending on its phase, but is typically > 3.0 eV.[21] Further increases in the band gap yielded no change in the simulation results.

c) Values in the range -4.0 eV to -4.3 eV have been reported for $TiO_2$.[22,23] We chose the value in this range could best account for the JV data and the Stabilise and Pulse measurements.

d) The mobility of electrons in $TiO_2$ varies depending upon both the phase and crystallinity of the material.[24] We have used 0.1 cm$^2$V$^{-1}$s$^{-1}$ as a conservative estimate from the values found in the literature.

e) The relative permittivity of $TiO_2$ depends upon both its phase and orientation. For films prepared using the sol-gel method, values in the range 19-64 have been recorded, depending upon the annealing temperature.[25] We have used 19 as this film was annealed at the temperature most similar to ours.

f) The effective mass of electrons in $TiO_2$ varies dramatically between different phases.[26] As the structure of the $TiO_2$ was unknown for these devices, we have assumed $m^* \approx m_e$ and calculated the density of states expected in the parabolic band approximation for this value of $m^*$. An under (over) estimate of this parameter would results in an over (under) estimate of the surface recombination velocity at the perovskite/$TiO_2$ interface as the rate of interfacial recombination increases as the density of states in the transport layer increases.

g) The majority carrier surface recombination velocity was observed to have a large effect on the shape of the simulated Stabilise and Pulse curves at low bias voltages. Thus, it's value was varied to match the experimental results, with larger values found to lead to less 's-shaped' JV curves at low bias voltages.

h) Value calculated from the supplier webpage is 3.16 eV. However, this parameter was set to 3 eV as further increases in the band gap yielded no change in the simulation results, but increased the time taken per simulation.

i) Values in range of -5.0 eV to -5.2 eV have been reported for Spiro-OMeTAD.[20,27] We chose the value in this range could best account for the JV data and the Stabilise and Pulse measurements.

j) Values chosen to yield a conductivity of ~$7 \times 10^{-5}$ Scm$^{-1}$, within the range of values found for Spiro-OMeTAD samples doped in a similar manner to those in this work.[28]

k) Typical value for organic semiconductors.

l) No value could be found in the literature for Spiro-OMeTAD so we have used $10^{20}$ cm$^{-3}$ as this lies within the range of values commonly found for organic semiconductors.[11]

**Supplementary Note One**

In this Supplementary Note, we consider the situation where the energetic offset to the transport layers, $\Delta E_{TL}$, is not the same at the perovskite/HTL and perovskite/ETL interfaces. For concreteness, we will consider the case where the ETL has a fixed $\Delta E_{TL}$ of 0.15 eV and we vary $\Delta E_{TL}$ for the HTL as described in the main text. We chose this fixed value of $\Delta E_{TL}$ so that we could explore the cases where a given interface has both a larger and smaller energetic offset to the perovskite in a single parameter sweep. Additionally, we increased the surface recombination velocity to $100\,\text{cm s}^{-1}$, such that the recombination losses around $V_{OC}$ are always dominated by surface recombination. We did this for two reasons: first, the case where bulk recombination dominates has already been discussed [29,30] and secondly, the model described in the main text is largely applicable to devices where surface recombination losses dominate. Furthermore, we use the doped inorganic parameter set to carry out these simulations to remove the additional complication of there being large voltage drops in the transport layers.

The results of our simulations for asymmetric $\Delta E_{TL}$ are shown in **Figure S20e**. First, we will discuss the novel phenomena which are observed under these conditions, and then we will consider how the model outlined in the main text must be adjusted to describe transport layer asymmetry and the validity of our experimental method under these conditions. The asymmetry of the transport layers' energetics results in an electronic carrier imbalance in the perovskite bulk where the nature of the carrier in excess (i.e., electron or hole) generally depends upon which transport layer has a smaller value of $\Delta E_{TL}$. For example, in the case where the HTL is perfectly aligned with the perovskite valence band, there will be an excess of holes in the bulk of the perovskite, making it effectively p-type (see inset of **Figure S20a**). This behaviour occurs irrespective of the presence of a mobile ionic species, and generally results in worse overall device performance due to the reduced conductivity of the minority species, though an increase in $V_{OC}$ can also be observed for devices limited by bulk recombination as the rate of this process is now determined by the minority carrier density.[29]

When mobile ions are present, the effective doping of the perovskite results in a non-zero excess ion density in the perovskite bulk (see **Figure S20a**). As our simulations assume that all the ionic charge is confined to the perovskite layer and that ionic charge is conserved, this results in the interfacial ionic accumulation and depletion regions containing unequal amounts of charge. Thus, the ionic accumulation/depletion regions on the two sides of the device no longer invert at the same applied bias (see Figures **S20c-d**). This means that there is no longer a single potential at which the ion distribution is uniform (i.e., $V_{flat}$), but instead two "inversion voltages". We have labelled these as $V_{HTL}$ and $V_{ETL}$, respectively on **Figures S20c-e**. We note that these voltages were found to be almost independent of the mobile ion density (down to a mobile ion density of $10^{15}\,\text{cm}^{-3}$). Additionally, in **Figure S20e**, we have drawn on the difference in the equilibrium values of the transport layers' Fermi energies, $\Delta E_{F,TLs}$ which is what determines the built-in potential for symmetric devices with highly doped transport layers (see main text). Interestingly, $V_{ETL}$ and $V_{HTL}$ are symmetrically distributed around $\Delta E_{F,TLs}$, with the offset decreasing as $\Delta E_{F,TLs}$ decreases. These facts will be of relevance when considering how asymmetry of the transport layers' energetics could affect the interpretation of our experimental data.

Considering these theoretical insights, we believe that the framework described in the main text is still applicable to a device with asymmetric contacts and where losses are dominated by surface recombination. However, in the asymmetric case, the effect of ions at each interface must be considered separately as the presence of ions will only reduce the minority carrier population at an interface compared to the case with no ions after the inversion voltage for that interface has been reached. Relative to a symmetric device with the same $\Delta E_{F,TLs}$, the inversion voltage will be below $V_{flat}$ at the interface to the transport layer with the worse energetic alignment to the perovskite, and that at the opposite interface will be above $V_{flat}$. In the region between the two inversion voltages, surface recombination will only be supressed relative to the case with no mobile ions at one interface, namely the one with worse energetic alignment to the perovskite. However, the voltage regimes above and below this intermediate region are analogous to the voltage being above and below $V_{flat}$ in the symmetric case, respectively. This can be seen in our results as, for $\Delta E_{HTL} < 0.15$ eV, both $V_{ETL}$ and $V_{HTL}$ are above $V_{OC}$, meaning that the presence of mobile ions does not increase its value. Conversely, for $\Delta E_{HTL} > 0.15$ eV, both $V_{ETL}$ and $V_{HTL}$ are below $V_{OC}$, and thus $V_{OC}$ increases in the presence of a mobile ionic species. The increase in $V_{OC}$ due to mobile ions is not as large as is found in the symmetric case (see **Figure 2b**) as the effective doping of the perovskite layer due to the asymmetric contacts means that the distribution of the ionic charge no longer dominates the device's electrostatics. Thus, there is a smaller difference between the behaviour of the devices with and without ions.

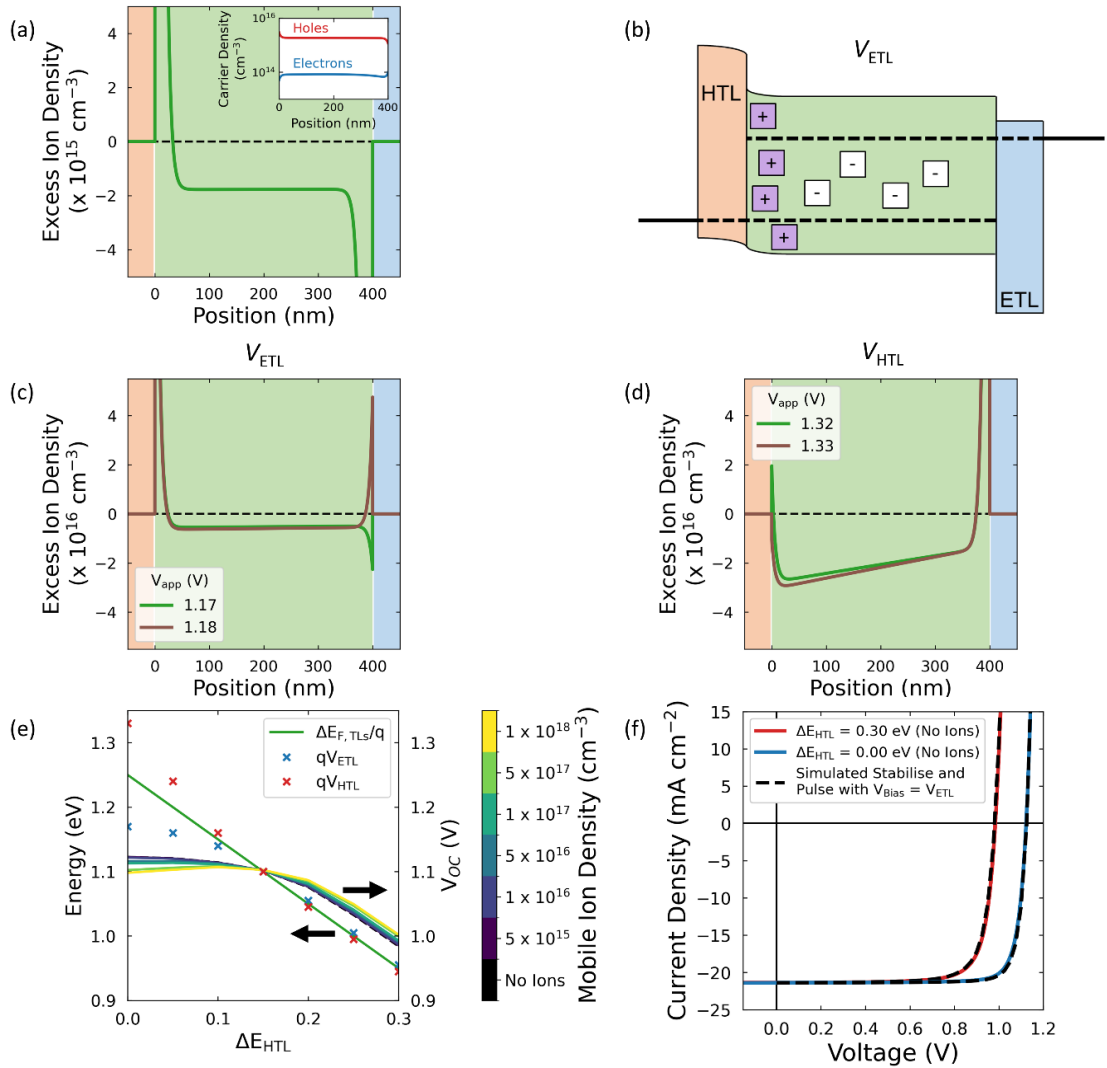

**Figure S20**: (a) Ionic and electronic carrier distributions at $V_{OC}$ in the case where $\Delta E_{TL}$ is 0.15 eV at the perovskite/ETL interface and there is no offset at the perovskite/HTL interface. (b) Schematic band diagram at the voltage at which the ion population inverts at the perovskite/ETL interface ($V_{ETL}$) in the case where $\Delta E_{TL}$ is 0.15 eV at the perovskite/ETL interface and there is no offset at the perovskite/HTL interface. Inversion of the ionic population at (c) the perovskite/ETL interface and (d) the perovskite/HTL interface for the same device as shown in (a). Ionic inversion occurs at a lower applied voltage at the perovskite/ETL interface due to the excess hole density in the perovskite bulk. (e) $V_{OC}$ as a function of ion density and energetic offset to the HTL ($\Delta E_{HTL}$) for a fixed energetic offset to the ETL of 0.15 eV. We have also marked the difference in the equilibrium Fermi levels of the transport layers, ($\Delta E_{F,TLs}$) and the inversion voltages at the perovskite/ETL and perovskite/HTL interfaces ($V_{ETL}$ and $V_{HTL}$, respectively). (f) Validation of the Stabilise and Pulse (SaP) measurement in the case of asymmetric energetic offsets to the transport layers. The parameters used are the same as those used to make figure (e) and, when ions have been included, we have used an ion density of $10^{18}$ cm⁻³. The coloured lines show the JV curves obtained with the value of $\Delta E_{HTL}$ indicated in the legend and in the absence of mobile ions. The black dashed lines indicated the result of a simulated Stabilise and Pulse measurement performed at a prebias voltage equal to the inversion voltage at the interface with the worse energetic alignment (1.18 V for $\Delta E_{HTL} = 0.00$eV and 0.94 V for $\Delta E_{HTL} = 0.30$ eV). These agree well with the JVs obtained in the absence of mobile ions, showing that a Stabilise and Pulse measurement can still be used to probe the impact of ions on $V_{OC}$ in the case of asymmetric transport layer energetics.

Lastly, we consider the impact these findings have on our experimental data. If there is no single $V_{flat}$ in the case where the transport layers have asymmetric energetic alignments, it is not obvious that *any* prebias potential will result in a pulsed JV which reproduces that of the ion-free device. Fortunately, our simulations suggest that the Stabilise and Pulse method is still viable as, in situations where there is an asymmetry in $\Delta E_{TL}$, surface recombination is dominated by the interface with the worse energetic alignment (i.e., larger $\Delta E_{TL}$ value). Thus, in a device whose recombination current is largely limited by surface recombination, what matters in terms of the JV performance is the minority carrier density at the interface with the larger energetic offset between perovskite and the transport layer. This can be matched to the case with no mobile ions by applying a voltage prebias which corresponds to the flat ion potential on the side of the device with the larger $\Delta E_{TL}$. In this case, our simulations suggest that the pulsed JV measurement will match the JV of an equivalent device with no mobile ions (see **Figure S20f**). As discussed above, the interface with the larger energetic offset will always have the lower inversion voltage and this will always be below $\Delta E_{F,TLs}$ meaning that it is highly likely this voltage will be included in the range of prebias values sampled in our measurements. Furthermore, in devices with larger energetic offsets to the transport layers (and hence lower values of $\Delta E_{F,TLs}$), our simulations suggest that there is not a large difference between $V_{ETL}$ and $V_{HTL}$, implying that it may not be a bad approximation to assume a single $V_{flat}$ in this situation.

**Supplementary Note Two**

In this Supplementary Note, we derive **Equation 1** in the main text, which gives an analytic expression for the change in electrostatic potential across the transport layers at the flat band condition in the perovskite, $V_{flat}$. In this derivation, the transport layers are treated as perfectly symmetric, undoped semiconductors, and we illustrate the geometry used in **Figure S21a**. In this figure, we have taken advantage of the fact that the electric field, $E$, at the perovskite/transport layer interfaces must be zero when we are at the flat band condition in the perovskite. This allows us to remove the perovskite layer from our consideration as its removal does not affect the boundary conditions for $V(x)$ at the newly formed ETL/HTL interface. By considering this diagram, we see that we can express the hole carrier density at position $x$ in the hole transport layer (HTL) as

$$p(x) = N_V \exp\left(\frac{E_V - \varphi - qV(x)}{k_BT}\right) \equiv n_0 \exp\left(\frac{-eV(x)}{k_BT}\right) \tag{S1}$$

$q$ is the electron charge, $k_B$ is the Boltzmann constant, $T$ is the temperature, $N_V$ is the density of states at the valence band edge, $E_V$ is the ionisation potential of the HTL, $\varphi$ is the work function of the electrode and $n_0$ the carrier density in the transport layer at the interface with the electrode, which is as defined via $n_0 = N_V \exp[(E_V - \varphi)/k_BT]$. We can substitute this expression into Poisson's equation to yield

$$\frac{d^2V(x)}{dx^2} = -\frac{q}{\varepsilon\varepsilon_0} n_0 \exp\left(\frac{-qV(x)}{k_BT}\right) \tag{S2}$$

$\varepsilon$ is the permittivity of free space and $\varepsilon_0$ is the relative permittivity of the transport layers. By integrating this equation with respect to $x$, we can find the electric field, $E$, in the HTL

$$E(x) = -\frac{dV(x)}{dx} = \pm\sqrt{\frac{2n_0q}{\varepsilon\varepsilon_0}\left(\frac{k_BT}{q}\exp\left(\frac{-eV(x)}{k_BT}\right) - C\right)} \tag{S3}$$

$C$ is a constant of integration, which can be found by using the following boundary conditions

$$E(x = w_{HTL}) = 0 \tag{S4}$$

$$V(x = w_{HTL}) = \frac{1}{2}\left(V_{BI} - V_{flat}\right) \equiv \frac{1}{2}\tilde{V} \tag{S5}$$

$$\Rightarrow C = \frac{k_BT}{q}\exp\left(\frac{-q\tilde{V}}{2k_BT}\right) \tag{S6}$$

The first boundary condition follows from the fact we are trying to find an expression for the potential at which we reach the flat band condition in the perovskite and the second follows from the symmetry of the transport materials, which means that half of the total change in voltage occurs over the HTL. Additionally, from **Figure S21a**, we can see that the gradient of $V(x)$ is positive for all $x$, implying a negative electric field (i.e., holes move to the left). Thus, we can discard the positive root in **Equation S3**. We note that, due to the square root in **Equation S3**, this solution is only valid for $V < \tilde{V}$.

Following this, **Equation S3** can be integrated again to find that

$$D - x = \frac{k_BT}{q}\sqrt{\frac{2\varepsilon\varepsilon_0}{n_0qC}}\arcsin\left(\sqrt{\frac{qC}{k_BT}}\exp\left(\frac{eV(x)}{2k_BT}\right)\right) \tag{S7}$$

$D$ is a second constant of integration, which can be found by using the boundary condition

$$V(x = 0) = 0 \tag{S8}$$

$$\Rightarrow D = \frac{k_BT}{q}\sqrt{\frac{2\varepsilon\varepsilon_0}{n_0qC}}\arcsin\left(\sqrt{\frac{qC}{k_BT}}\right) \tag{S9}$$

By combining **Equations S3, S6 and S9**, we can write that

$$x = \sqrt{\frac{2\varepsilon\varepsilon_0 k_B T}{n_0 q^2}} \exp\left(\frac{e\tilde{V}}{4k_B T}\right)\left[\arcsin\left(exp\left(\frac{q(V(x)-\frac{1}{2}\tilde{V})}{2k_B T}\right)\right) - \arcsin\left(\exp\left(\frac{-q\tilde{V}}{4k_B T}\right)\right)\right] \quad (S10)$$

$$\Rightarrow w_{HTL} = \sqrt{\frac{2\varepsilon\varepsilon_0 k_B T}{n_0 q^2}} \exp\left(\frac{e\tilde{V}}{4k_B T}\right)\left[\frac{\pi}{2} - \arcsin\left(\exp\left(\frac{-q\tilde{V}}{4k_B T}\right)\right)\right] \quad (S11)$$

as written in the main text. In **Figure S21b**, we plot the numerical solution for $\tilde{V}$ as a function of the energetic offset between the electrode work function and the HTL valance band using the transport layer parameters from the doped organic parameter set (see **Table S1**).

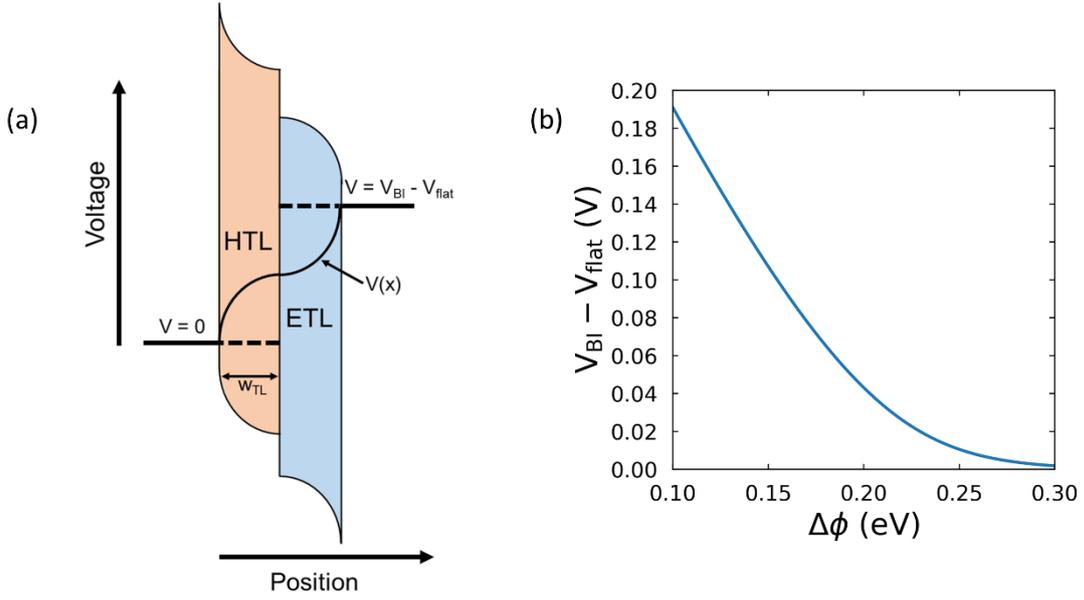

**Figure S21**: (a) Diagram of the geometry used to derive **Equation S10**. (b) Value of $\tilde{V} = V_{BI} - V_{flat}$ found using **Equation S11** as a function of the energetic offset between the work function of the electrode and the conduction (valance) band of the ETL (HTL), $\Delta\varphi$. This parameter controls the majority carrier density at the electrode/transport layer interfaces, $N_0$. The range of $\Delta\varphi$ shown here cover the range used in our simulation results, and the calculated offset between $V_{BI}$ and $V_{flat}$ agrees with that found using the full Driftfusion simulation (compare to **Figure S2**).

**Supplementary Note Three**

The simulations whose results are shown in **Figure 1** and **Figure 4** of the main text were performed on highly idealised perovskite devices. Thus, to verify that we would expect to see the same trends in real devices, we performed simulations using parameters representative of the Au/Spiro-OMeTAD/MAPbI$_3$/(C$_{60}$-BA)/TiO$_2$/FTO device stack (parameters given in **Table S6**). We modelled the effect of the C$_{60}$-BA by reducing the surface recombination velocity of holes at the perovskite/TiO$_2$ interface by a factor of one thousand. Although this may seem to be a simplistic treatment of the C$_{60}$-BA, we find that this change allowed us to recreate the observed trends in the Stabilise and Pulse measurements and the impact of mobile ions on $V_{OC}$, as is shown in **Figure S22**. JV parameters with and without mobile ions are given in **Table S7**, below.

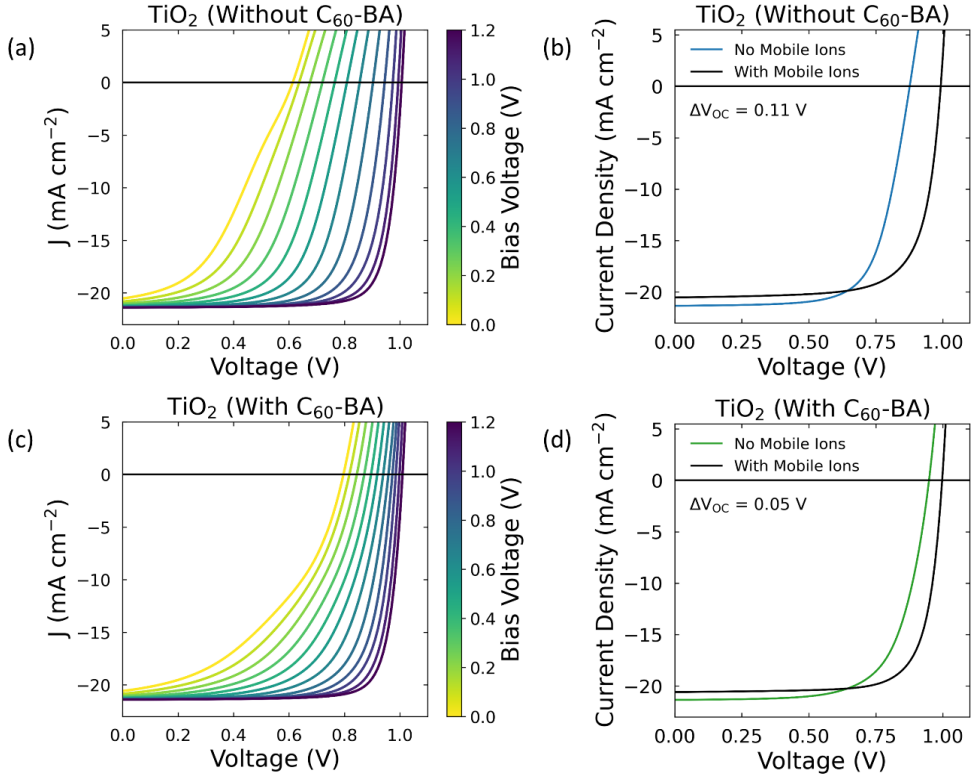

**Figure S22:** Simulated Stabilise and Pulse measurements (left-hand column) and JV curves with and without the inclusion of a mobile ionic species (right-hand column) for (a-b) devices without C$_{60}$-BA and (c-d) devices with C$_{60}$-BA. Comparing this figure to **Figure 3** in the main text, we see that the simulations can accurately reproduce the experimental trends.

| Device | With or Without Ions | J$_{SC}$ (mA cm$^{-2}$) | V$_{OC}$ (V) | FF | PCE (%) |
|---|---|---|---|---|---|
| Without C$_{60}$-BA | With | 20.50 | 0.99 | 0.73 | 14.9 |
| | Without | 21.30 | 0.88 | 0.70 | 13.0 |
| With C$_{60}$-BA | With | 20.60 | 1.00 | 0.76 | 15.7 |
| | Without | 21.20 | 0.95 | 0.69 | 13.9 |

**Table S7:** JV parameters from the JV curves shown in **Figure S21b** (without C$_{60}$-BA) and **Figure S21d** (with C$_{60}$-BA).

**Supplementary Note Four**

In this Supplementary Note, we summarise our Stabilise and Pulse measurements on Au/Spiro-OMeTAD/MAPbI$_3$/(C$_{60}$-BA)/SnO$_2$/FTO devices. As for the TiO$_2$ devices, we summarise the JV data in **Tables S8-9** and **Figure S23** (below). Additionally, on **pages 29-30**, we provide plots of the full Stabilise and Pulse measurements, dJ/dV|$_{V=V_{OC}}$ versus $V_{bias}$ and the quasi-steady state (QSS) and stabilise and pulse (SaP) JVs evaluated at $V_{flat}$. Plots of the full Stabilise and Pulse measurements include the polynomial fits used to evaluate dJ/dV|$_{V=V_{OC}}$ (see **Methods**). Data is shown up to the highest $V_{bias}$ necessary to determine the QSS $V_{OC}$.

**Table S8:** Summary of the JV parameters for the Au/Spiro-OMeTAD/MAPbI$_3$/SnO$_2$/FTO devices. JV parameters are given for three cases: as measured under the Solar Simulator using the protocol described in the **Methods** section (parameters taken from the reverse scan), as measured with the mobile ions at quasi-steady state (QSS JV) and as measured using Stabilise and Pulse measurements carried out at $V_{bias}$ = $V_{flat}$ (SaP JV). Additionally, we provide the value of $V_{flat}$ extracted from our analysis of the Stabilise and Pulse data in column 2.

| Device Number | $V_{flat}$ (V) | Method | $J_{SC}$ (mA cm$^{-2}$) | $V_{OC}$ (V) | FF | PCE (%) |
|---|---|---|---|---|---|---|
| 1 | 0.68 ± 0.05 | Solar Sim | 21.30 | 1.053 | 0.75 | 16.82 |
| | | QSS JV | 19.45 | 0.956 | 0.54 | 10.08 |
| | | SaP JV | 21.77 | 0.826 | 0.64 | 11.47 |
| 2 | 0.68 ± 0.05 | Solar Sim | 21.50 | 1.054 | 0.71 | 16.11 |
| | | QSS JV | 21.03 | 0.990 | 0.59 | 12.22 |
| | | SaP JV | 21.45 | 0.848 | 0.66 | 12.03 |
| 3 | 0.67 ± 0.05 | Solar Sim | 21.49 | 1.031 | 0.73 | 16.09 |
| | | QSS JV | 17.23 | 0.969 | 0.65 | 10.85 |
| | | SaP JV | 21.38 | 0.789 | 0.62 | 10.46 |
| 4 | 0.65 ± 0.05 | Solar Sim | 20.51 | 1.033 | 0.73 | 15.51 |
| | | QSS JV | 17.80 | 0.972 | 0.58 | 10.00 |
| | | SaP JV | 20.20 | 0.771 | 0.62 | 9.59 |

**Table S9:** Summary of the JV parameters for the Au/Spiro-OMeTAD/MAPbI$_3$/C$_{60}$-BA/SnO$_2$/FTO devices. For further details, see the caption of **Table S8**.

| Device Number | $V_{flat}$ (V) | Method | $J_{SC}$ (mA cm$^{-2}$) | $V_{OC}$ (V) | FF | PCE (%) |
|---|---|---|---|---|---|---|
| 1 | 0.73 ± 0.05 | Solar Sim | 20.10 | 1.095 | 0.75 | 16.60 |
| | | QSS JV | 19.24 | 1.042 | 0.75 | 14.96 |
| | | SaP JV | 20.39 | 0.948 | 0.71 | 13.73 |
| 2 | 0.73 ± 0.05 | Solar Sim | 22.27 | 1.070 | 0.69 | 16.37 |
| | | QSS JV | 20.61 | 0.977 | 0.72 | 14.42 |
| | | SaP JV | 23.05 | 0.940 | 0.71 | 13.73 |
| 3 | 0.70 ± 0.05 | Solar Sim | 20.85 | 1.104 | 0.73 | 16.86 |
| | | QSS JV | 20.75 | 1.012 | 0.74 | 15.57 |
| | | SaP JV | 21.09 | 0.979 | 0.67 | 13.91 |
| 4 | 0.72 ± 0.05 | Solar Sim | 21.36 | 1.101 | 0.72 | 16.85 |
| | | QSS JV | 21.08 | 1.029 | 0.77 | 16.70 |
| | | SaP JV | 21.75 | 0.989 | 0.67 | 14.49 |

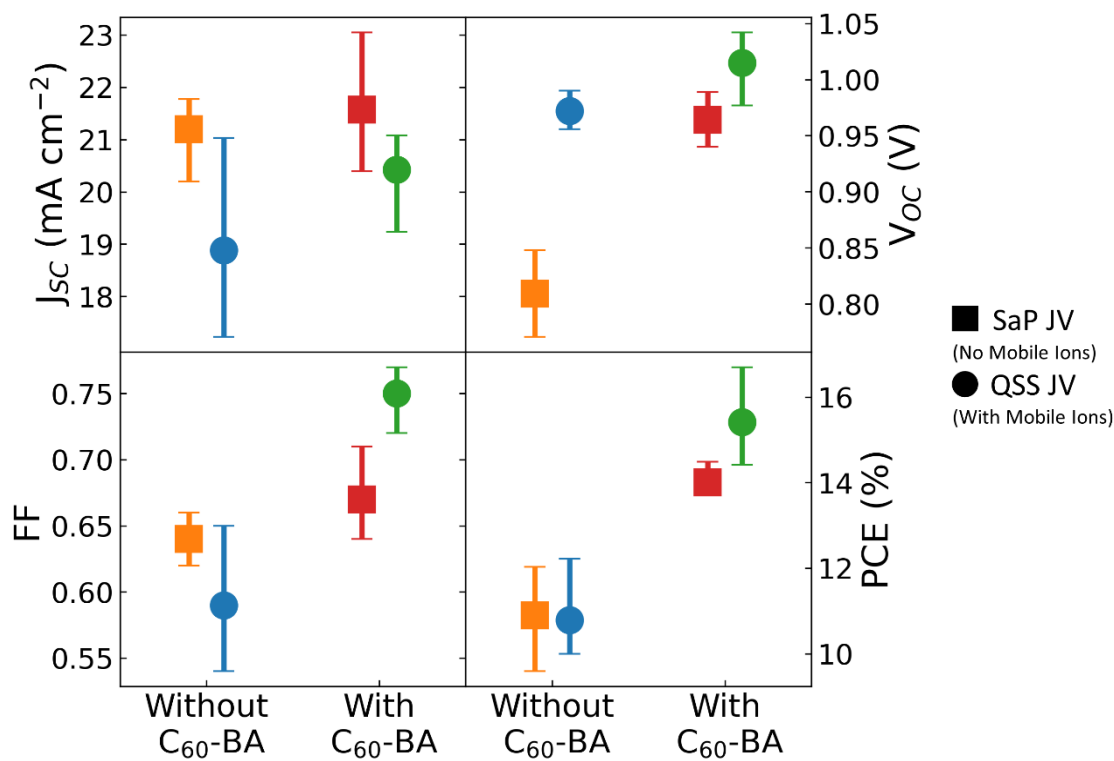

**Figure S23:** Summary of the JV parameters extracted from the Stabilise and Pulse (SaP) JVs and quasi-steady state (QSS) JVs for all measured SnO$_2$ devices. Error bars indicate the range of measured values. Considering the QSS JVs, we see that the inclusion of C$_{60}$-BA increases PCE and this is largely due to increases in $J_{SC}$ and FF, as was observed in the TiO$_2$ devices (see **Figure 2e**). Furthermore, the devices using SnO$_2$ show the same trends in $V_{OC}$ as the TiO$_2$ devices, namely that ions increase $V_{OC}$ with and without the inclusion of C$_{60}$-BA, and the size of this improvement decreases when the C$_{60}$-BA is present. Additionally, in the devices with C$_{60}$-BA, we find that the presence of mobile ions increases PCE, as found in the TiO$_2$ devices. However, the devices without C$_{60}$-BA don't follow this trend, largely due to their low FF. This contrasts to what was observed in the TiO$_2$ devices, where the QSS JVs had higher FFs than the SaP JVs. By comparing the QSS JVs of the SnO$_2$ and TiO$_2$ devices without C$_{60}$-BA, we see that this difference originates from the rapid loss in photocurrent observed at low voltages in the SnO$_2$ devices. Since this behaviour is less pronounced in the devices with C$_{60}$-BA, we believe it is due to the interaction of the perovskite and SnO$_2$ layers. This highlights how well passivated interfaces are necessary for mobile ions to be beneficial to device performance.

## SnO$_2$ (Without C$_{60}$-BA) – <u>Device 1</u>

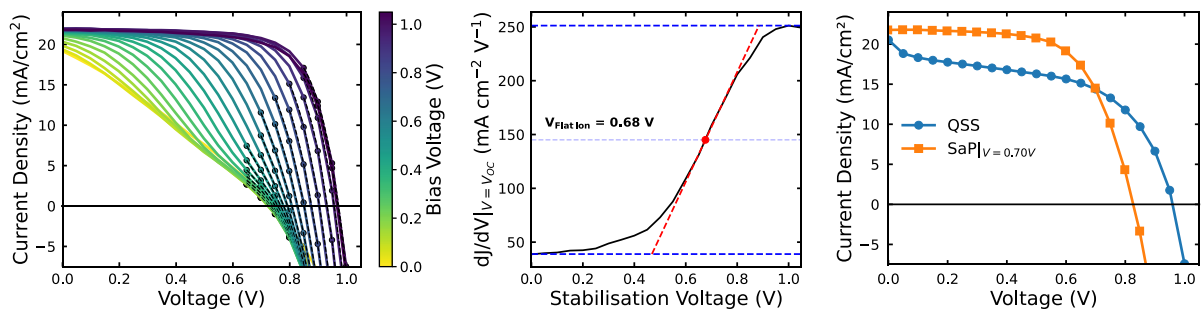

## SnO$_2$ (Without C$_{60}$-BA) – <u>Device 2</u>

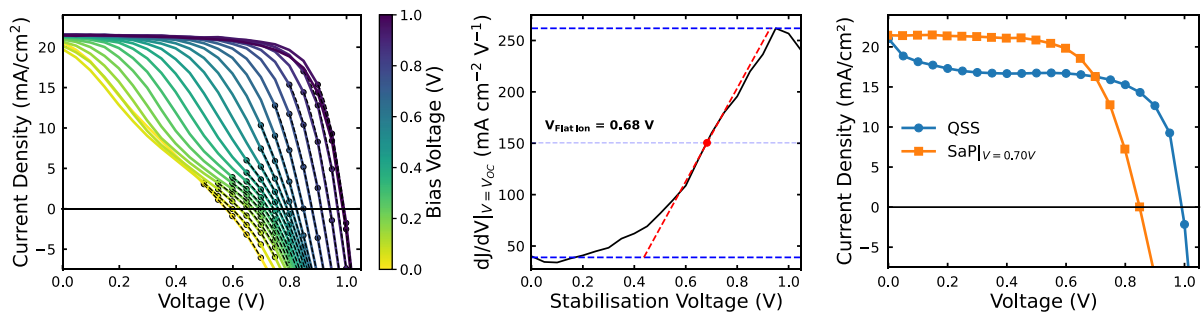

## SnO$_2$ (Without C$_{60}$-BA) – <u>Device 3</u>

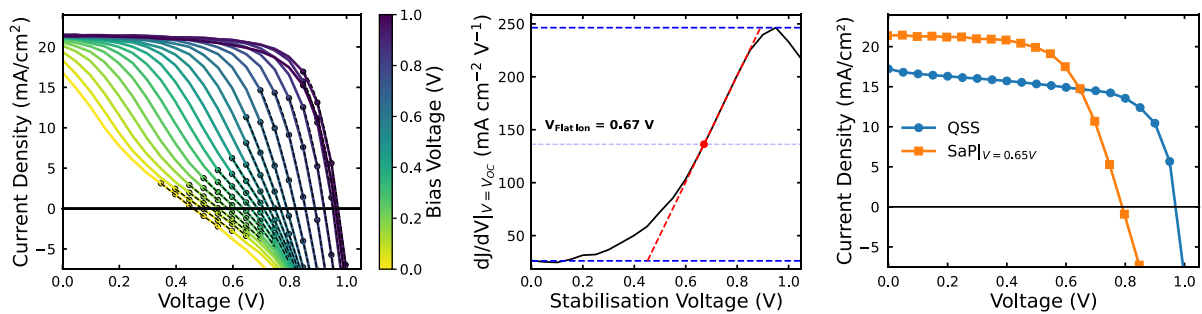

## SnO$_2$ (Without C$_{60}$-BA) – <u>Device 4</u>

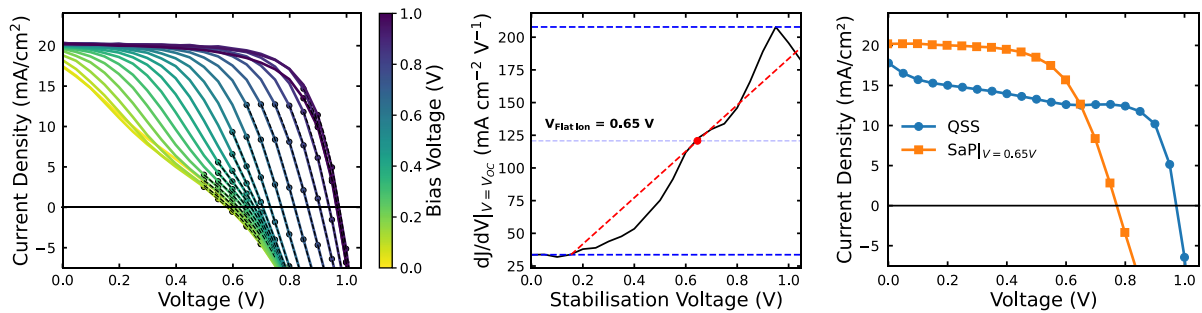

### SnO$_2$ (With C$_{60}$-BA) – <u>Device 1</u>

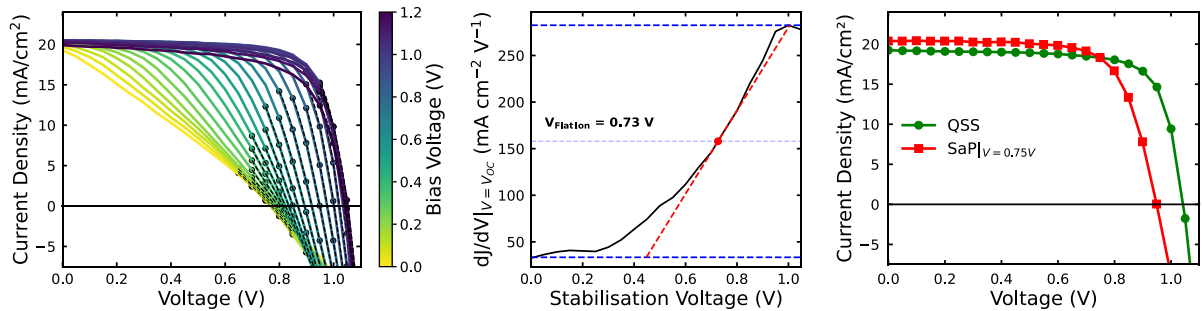

### SnO$_2$ (With C$_{60}$-BA) – <u>Device 2</u>

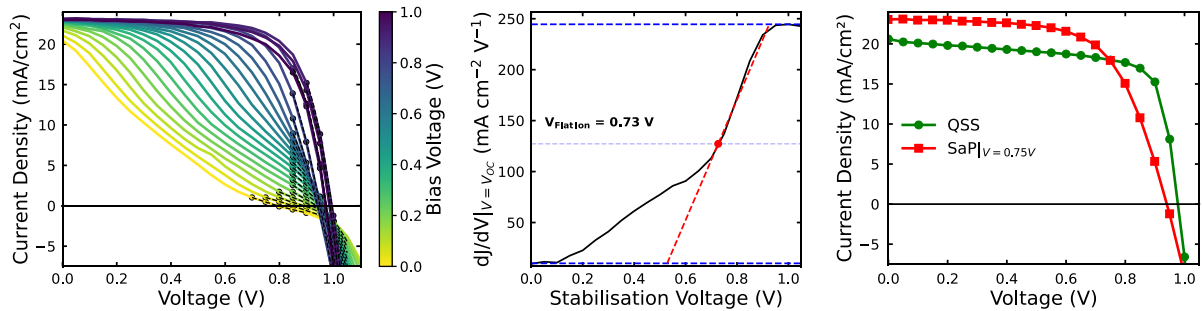

### SnO$_2$ (With C$_{60}$-BA) – <u>Device 3</u>

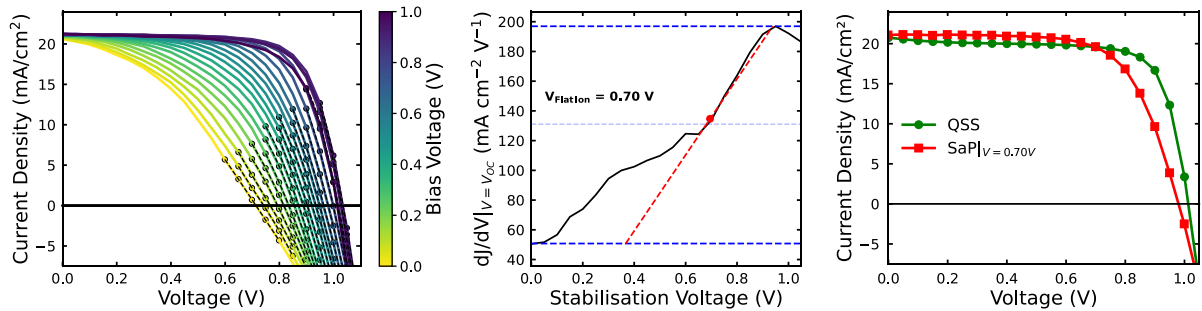

### SnO$_2$ (With C$_{60}$-BA) – <u>Device 4</u>

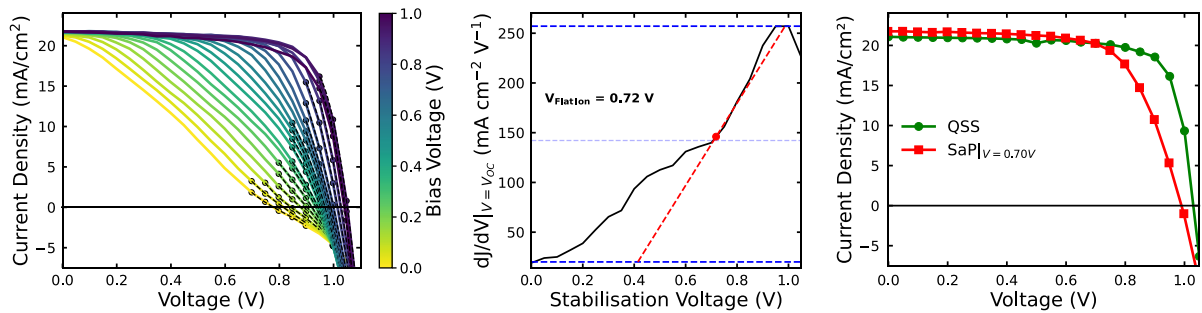

## Supplementary Note Five

In this Supplementary Note, we provide additional data regarding the Stabilise and Pulse measurements carried out on the Au/Spiro-OMeTAD/MAPbI$_3$/(C$_{60}$-BA)/TiO$_2$/FTO devices which were not included in the main text. This data includes plots of the full Stabilise and Pulse measurements, dJ/dV|$_{V=Voc}$ versus $V_{bias}$ and the quasi-steady state (QSS) and Stabilise and Pulse (SaP) JVs evaluated at $V_{flat}$. Plots of the full Stabilise and Pulse measurements include the polynomial fits used to evaluate dJ/dV|$_{V=Voc}$ (see **Methods**). Data is shown up to the highest $V_{bias}$ necessary to determine the QSS $V_{OC}$.

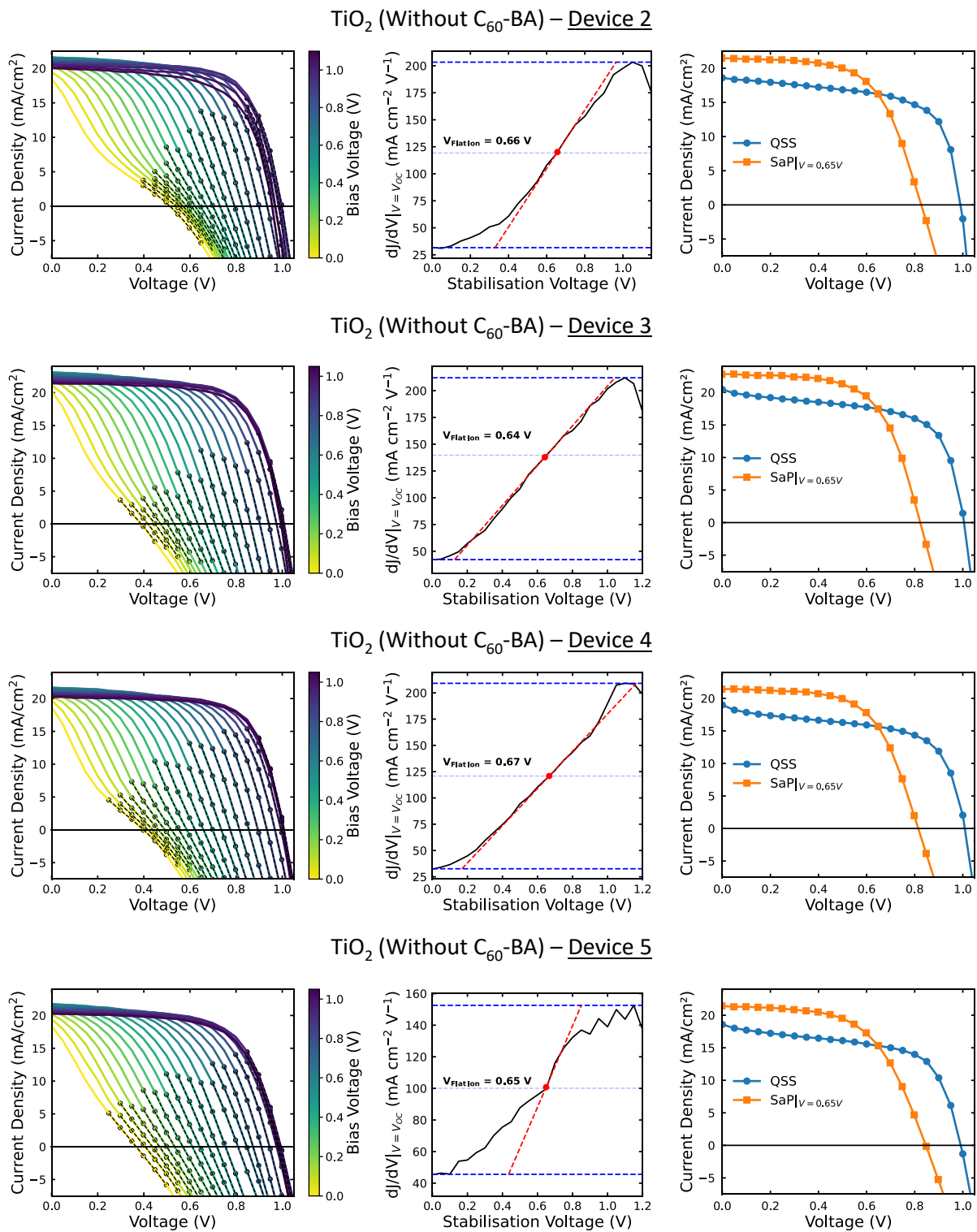

### TiO$_2$ (With C$_{60}$-BA) – <u>Device 1</u>

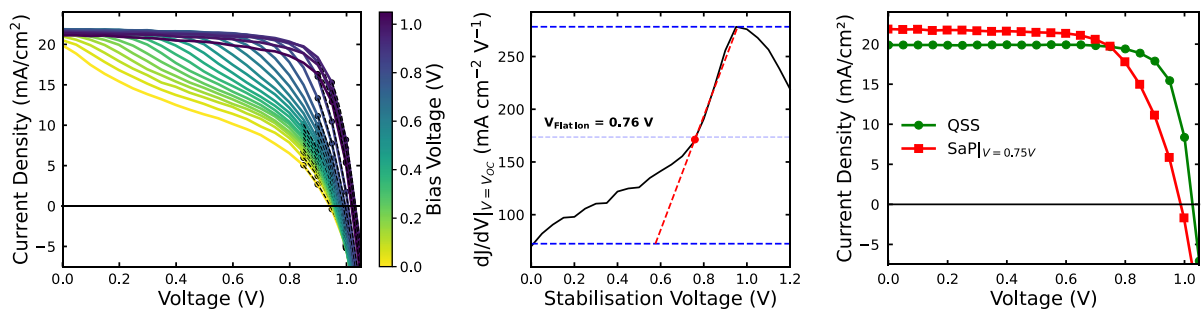

### TiO$_2$ (With C$_{60}$-BA) – <u>Device 2</u>

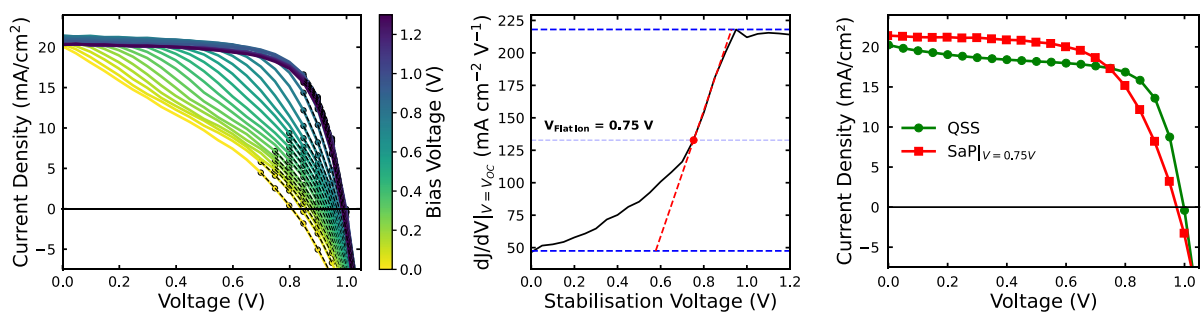

### TiO$_2$ (With C$_{60}$-BA) – <u>Device 3</u>

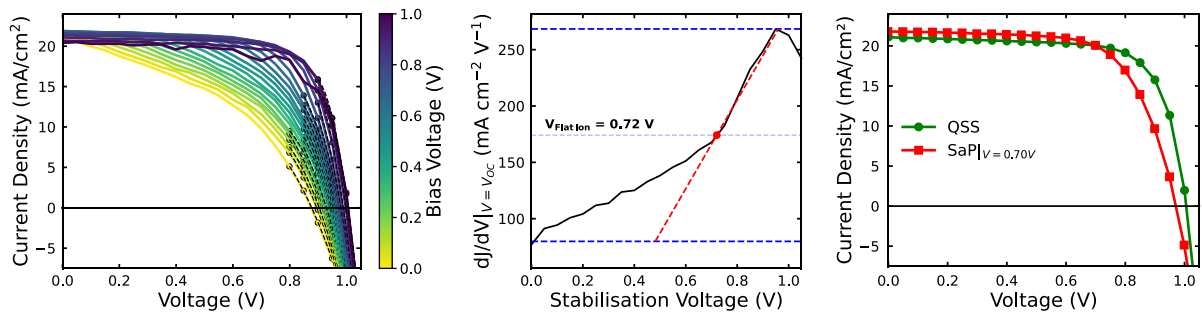

## TiO$_2$ (With C$_{60}$-BA) – <u>Device 4</u>

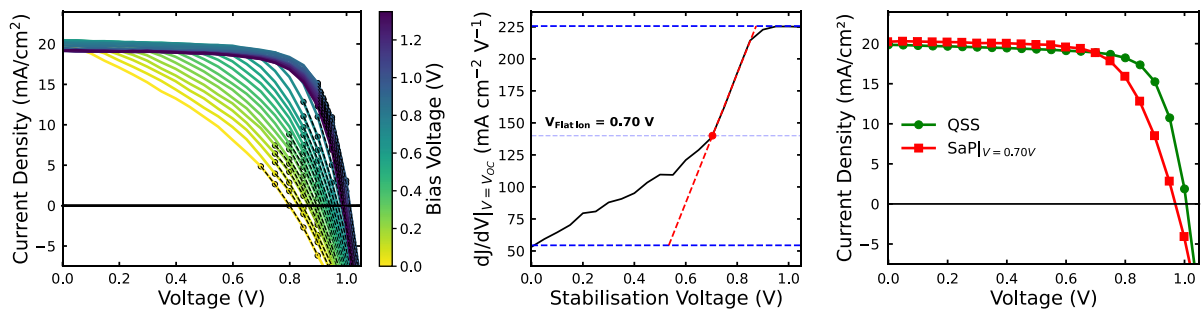

## TiO$_2$ (With C$_{60}$-BA) – <u>Device 6</u>

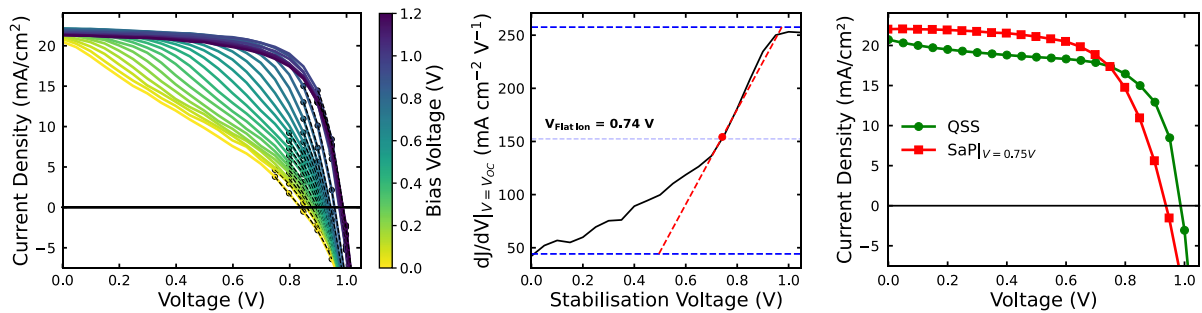